\documentclass[longauth]{aa}
\UseRawInputEncoding 
\usepackage[utf8]{inputenc}
\usepackage{pdflscape}
\usepackage{xcolor}
\usepackage{lscape}
 \usepackage[hidelinks]{hyperref}
   \hypersetup{
      colorlinks = true,
      citecolor ={cyan}
   }

\usepackage{graphicx}
\newcommand{\ergs}{${\rm erg}\,{\rm s}^{-1}$}
\newcommand{\msun}{$M_\odot$}
\newcommand{\kms}{km\,${\rm s}^{-1}$}
\newcommand{\smy}{$M_\odot\,{\rm yr}^{-1}$}
\newcommand{\starname}{R\,144}

\newcommand{\HeII}{He\,{\sc ii}\,$\lambda 4686$,}

\newcommand{\PVres}{P\,{\sc v}\,$\lambda \lambda 1118, 1128$}
\newcommand{\NVred}{N\,{\sc v}\,$\lambda 4945$}
\newcommand{\NVblue}{N\,{\sc v} $\lambda \lambda 4604, 4620$}
\newcommand{\NIII}{N\,{\sc iii}\,$\lambda \lambda 4634, 4641$}
\newcommand{\NIV}{N\,{\sc iv}\,$\lambda 4058$}
\usepackage{txfonts}
\begin{document}

   \title{The Tarantula Massive Binary Monitoring\\ V.  R\,144 --  a wind-eclipsing binary with a total mass $\gtrsim 140$\,\msun \thanks{Based on observations collected at the European Southern Observatory under program IDs 085.D-0704(A), 085.D-0704(B), 086.D-0446(A), 086.D-0446(B), 087.C-0442(A), 087.D-0946, 090.D-0212(A), 090.D-0323(A), 092.D-0136(A), 292.D-5016(A).}}
%   \author{T.\ Shenar,
%   H.\ Sana,
%   B.\ Pablo,    
%   P.\ Marchant,
%   J   
%   K.\ Dsilva,
%   TMBM collab.\
%   }

   \author{T.\ Shenar\inst{1}
          \and H.\ Sana\inst{1} 
          \and P.\ Marchant\inst{1}  
          \and B.\ Pablo\inst{2}             
          \and N.\ Richardson\inst{3}  
          \and A.\ F.\ J.\ Moffat\inst{4}
          \and T.\ Van Reeth\inst{1}          
        %   \and L.\ A.\ Almeida\inst{12,13}   
        %   \and R.\ Barb\'a\inst{14}      
%Start of alphabetic
         \and R.\ H.\ Barb\'a\inst{5}  
         \and D.\ M.\ Bowman\inst{1}  
         \and P.\ Broos\inst{6}  
        %   \and A.\ Z.\ Bonanos\inst{8}  
          \and P.\ A.\ Crowther\inst{7}     
          \and J.\ S.\ Clark\inst{8}\thanks{We are saddened to report the death of Dr Simon Clark, who passed away during the preparation of the manuscript.}
          \and A.\ de Koter\inst{9}             
          \and S.\ E.\ de Mink\inst{10, 9, 11}                 
          \and K.\ Dsilva\inst{1}    
          \and G.\ Gr\"afener\inst{12}              
          \and I.\ D.\ Howarth\inst{13}
        %   \and A.\ de Koter\inst{4,9}               
        %   \and S.\ E.\ de Mink\inst{9}    
        %   \and A.\ Herrero\inst{15}          
          \and N.\ Langer\inst{12}    
          \and L.\ Mahy\inst{1, 14}       
         \and J.\ Ma\'iz Apell\'aniz\inst{15}         
        %   \and A.\ F.\ J.\ Moffat\inst{5}
          \and A.\ M.\ T.\ Pollock\inst{7}    
        %   \and F.\ Tramper\inst{6}   
         \and F.\ R.\ N.\ Schneider\inst{16, 17}            
         \and L.\ Townsley\inst{6}      
         \and J.\ S.\ Vink\inst{18}             
        %   \and J.\ S.\ Vink\inst{10}     
          }

   \institute{\inst{1}{Institute of Astronomy, KU Leuven, Celestijnenlaan 200D, 3001 Leuven, Belgium}; \email{tomer.shenar@kuleuven.be} \\     
             \inst{2}{American Association of Variable Star Observers, 49 Bay State Road,
Cambridge, MA 02138, USA}\\
             \inst{3}{Department of Physics and Astronomy, Embry-Riddle Aeronautical University, 3700 Willow Creek Road, Prescott, AZ 86301, USA}\\
              \inst{4}{D\'epartement de physique and Centre de Recherche en Astrophysique 
                du Qu\'ebec (CRAQ), Universit\'e de Montr\'eal, C.P. 6128, Succ.~Centre-Ville, Montr\'eal, Qu\'ebec, H3C 3J7, Canada}\\      
              \inst{5}{Departamento de Astronom\'ıa, Universidad de La Serena, Av. Juan Cisternas 1200 Norte, La Serena, Chile}\\
              \inst{6}{Department of Astronomy \& Astrophysics, 525 Davey Laboratory, Pennsylvania State University, University Park, PA 16802, USA}\\
              \inst{7}{Department of Physics \& Astronomy, University of Sheffield, Hounsfield Road, Sheffield, S3 7RH, UK}\\
              \inst{8}{School of Physical Sciences, The Open University, Milton Keynes, UK}\\          
              \inst{9}{Anton Pannekoek Institute for Astronomy, University of Amsterdam, Postbus 94249, 1090 GE Amsterdam, The Netherlands }\\        
              \inst{10}{Max-Planck-Institut für Astrophysik, Karl-Schwarzschild-Straße 1, 85741 Garching, Germany}\\    
              \inst{11}{Harvard-Smithsonian Center for Astrophysics,
              60 Garden St., Cambridge, MA 02138, USA}\\     
              \inst{12}{Argelander-Institut fur Astronomie, Universit ¨ at Bonn, Auf dem H\"ugel 71, 53121 Bonn, Germany}\\
              \inst{13}{Department of Physics and Astronomy, University College London, Gower Street, London WC1E 6BT, UK}\\
              \inst{14}{Royal Observatory of Belgium, Avenue Circulaire 3, B-1180 Brussel, Belgium}\\              
              \inst{15}{Centro de Astrobiolog\'ia, CSIC-INTA, Campus ESAC, Camino bajo del castillo, E-28 692 Villanueva de la Canada, Madrid, Spain}\\       
              \inst{16}{Heidelberger Institut fur Theoretische Studien, Schloss-Wolfsbrunnenweg 35, 69118 Heidelberg, Germany} \\          
             \inst{17}{Astronomisches Rechen-Institut, Zentrum fur Astronomie der Universit\"at Heidelberg, M\"onchhofstr. 12-14, 69120 Heidelberg, Germany} \\              
              \inst{18}{Armagh Observatory and Planetarium, College Hill, Armagh, BT61 9DG, Northern Ireland}
              }

   \date{Received March 02, 2021; accepted April 06, 2021}

% \abstract{}{}{}{}{}
% 5 {} token are mandatory

  \abstract
   {The evolution of the most massive stars and their upper-mass limit remain insufficiently constrained. Very massive stars are characterized by powerful winds and spectroscopically appear as hydrogen-rich Wolf-Rayet (WR) stars on the main sequence. \starname~is the visually brightest WR star in the Large Magellanic Cloud (LMC). \starname~was reported to be a binary, making it  potentially the most massive binary thus observed. However, the orbit and properties of \starname~are  yet to be established.}
   {Our aim is to derive the physical, atmospheric, and orbital parameters of \starname~and interpret its evolutionary status.}
   {We perform a comprehensive spectral, photometric, orbital, and polarimetric analysis of \starname. Radial velocities are measured via cross-correlation. Spectral disentangling is performed using the shift-and-add technique. We use the Potsdam Wolf-Rayet (PoWR) code for the spectral analysis. We further present X-ray and optical light-curves of \starname, and analyse the latter using  a hybrid model combining wind eclipses and colliding winds to constrain the orbital inclination $i$.}
   {\starname~is an eccentric ($e=0.51$)  $74.2-$d binary comprising two relatively evolved (age $\approx 2\,$Myr), H-rich WR stars (surface mass fraction $X_{\rm H} \approx 0.4$). The hotter primary (WN5/6h, $T_* = 50\,$kK) and the cooler secondary (WN6/7h, $T_* = 45\,$kK) have nearly equal masses, with $M \sin^3 i = 48.3\pm1.8$\,\msun~and $45.5\pm1.9$\,\msun, respectively. The combination of low rotation and H-depletion observed in the system is well reproduced by contemporary evolution models that include boosted mass-loss at the upper-mass end. The systemic velocity of \starname~and its relative isolation suggest that it was ejected as a runaway from the neighboring \object{R\,136} cluster. The optical light-curve shows a clear orbital modulation that can be well explained as a combination of two processes: excess emission stemming from wind-wind collisions and double wind eclipses. Our light-curve model implies an orbital  inclination of $i = 60.4\pm1.5^\circ$, resulting in accurately constrained dynamical masses of $M_{\rm 1, dyn} = 74\pm4\,M_\odot$ and $M_{\rm 2, dyn} = 69\pm4\,M_\odot$. Assuming that both binary components are core H-burning, these masses are difficult to reconcile with the derived luminosities ($\log L_{1, 2}/L_\odot = 6.44, 6.39 $), which correspond to evolutionary masses of the order of $M_{\rm 1, ev} \approx 110\,M_\odot$ and $M_{\rm 2, ev} \approx 100\,M_\odot$. Taken at face value, our results imply that both stars have high classical Eddington factors of $\Gamma_{\rm e}=0.78\pm0.10$. If the stars are on the main sequence, their derived radii ($R_* \approx 25\,R_\odot$) suggest that they are only slightly inflated, even at this high Eddington factor. Alternatively, the stars could be core-He burning, strongly inflated from the regular size of  classical Wolf-Rayet stars ($\approx 1\,R_\odot$), a scenario that could help resolve the observed mass discrepancy.}
   {\starname~is one of the few very massive extragalactic binaries ever weighed without usage of evolution models, but poses several challenges in terms of the measured masses of its components. To advance, we strongly advocate for future polarimetric, photometric, and spectroscopic monitoring of \starname~and other very massive binaries. }

   \titlerunning{The Tarantula Massive Binary Monitoring: V. R144}
   \authorrunning{T. Shenar et al.}

   \maketitle
%
%-------------------------------------------------------------------
%
\section{Introduction}\label{sec:intro}

The upper-mass limit throughout cosmic times remains a fundamental uncertainty in models of star formation and feedback \citep{Figer2005}. Theoretical estimations of this parameter widely vary from $\approx 120$~\msun\ to a few thousands of solar masses, depending on the metallicity and modeling assumptions \citep[e.g.,][]{Larson1971, Oey2005}.
The most massive stars largely dictate the radiative and mechanical energy budget of their host galaxies \citep{Doran2013, Bestenlehner2014, Ramachandran2019}. They are invoked as chemical factories to explain the presence of multiple populations in globular clusters \citep{Gieles2018, Bastian2018, Vink2018}, and are the presumed progenitors of the most massive black holes (BH) observed with the LIGO-VIRGO detectors \citep[e.g.][]{Abbott2020MostMassive} and pair-instability supernovae \citep[][]{Fryer2001, Woosley2007, Marchant2019}.

Much effort has been dedicated to finding the most massive stars in the Local Group.  It was realised by \citet{deKoter1997} that stars with current masses in the excess of $M \approx 100$~\msun\ spectroscopically appear as Wolf-Rayet (WR) stars: hot stars with  emission-line dominated spectra that form in their powerful stellar wind. While the WR phase is typically associated with evolved, core He-burning massive stars (dubbed classical WR stars), very massive stars already retain a WR-like appearance on the zero-age main sequence (see also \citealt{Maiz2017} for stars just below this limit).  Being relatively hydrogen and nitrogen rich, they are usually classified as WNh stars.

In terms of stellar mass, the current record-breaker is the WN5h star  \object{R\,136a1} (alias \object{\mbox{BAT99}~108}), followed by the WN5h star \object{R136c} (alias \object{\mbox{BAT99}~112}), with reported masses in the range of 200 to 300~\msun\ \citep{Crowther2010, Bestenlehner2020}. These objects are found in the core of the Tarantula nebula in the sub-solar metallicity environment ($Z\approx 0.5\,Z_\odot$) of the Large Magellanic Cloud (LMC). Other reported examples include the LMC WN5h star \object{VFTS 682} \citep{Bestenlehner2014} and several very massive WNh and Of stars in the Galactic clusters Arches \citep[e.g.][]{Figer2002, Najarro2004, Martins2008, Lohr2018} and \object{NGC~3603} \citep{Crowther2010}. Masses for these stars were inferred through luminosity calibrations under the assumption that the stars are single. 
Given the difficulty in disproving stellar multiplicity, the question of whether  these most massive stars are truly single remains open. 

A more reliable way to measure the masses of stars is offered by binary systems. With the measurement of both radial-velocity (RV) amplitudes $K_1$ and $K_2$, the minimum masses $M_1 \sin^3 i$ and $M_2 \sin^3 i$ can be derived from Newtonian mechanics, noting the strong dependence on the orbital inclination $i$. If the inclination can be derived, the masses then follow.  This is particularly achievable in eclipsing binaries such as \object{WR 43a} (alias \object{NGC 3603-A1}),  which enabled a  Keplerian mass estimate of $M_1 = 116\pm31$~\msun\ for the primary component \citep{Schnurr2008}. The inclination can also be constrained by other means, such as polarimetry, as performed by \citet{Shenar2017b} for \object{R\,145} in the LMC ( \object{\mbox{BAT99}~119}), or through an interferometric orbit determination (e.g., \object{WR~137} and \object{WR~138} in the Galaxy, \citealt{Richardson2016}).

In cases where the inclination is not known, mass calibrations of one or both of the binary components to their spectral classes and/or luminosities can convert the minimum masses to actual masses. While the calibration makes this method model-dependent, performing this in a binary provides a much more solid basis for the mass estimate compared to single-star estimates. The reason is twofold. First, by resolving the orbit of the two components, the danger of contamination by additional stellar components, as is the case for apparently-single stars, decreases. Second, a calibration of one of the components automatically results in the mass of the second component, which allows to test the consistency of the calibration for both components simultaneously. Examples include the Galactic binaries \object{WR~21a} \citep{Tramper2016} and \object{WR~20a} \citep{Bonanos2004, Rauw2004}. This method has recently been applied by \citet{Tehrani2019} to the LMC WN5h+WN5h binary \object{Melnick~34} (alias \object{\mbox{BAT99}~116})  to establish the most massive stellar components ever weighed in a binary: $M_1 = 139^{+21}_{-18}$~\msun\ and $M_2 = 127\pm17$~\msun.

With a Smith visual magnitude of $\varv = 11.15\,$mag \citep{Breysacher1999}, \object{\starname} (\emph{a.k.a.} \object{RMC~144}, \object{\mbox{BAT99}~118}, \object{Brey~89}, \object{HD~38282}) is the visually brightest WR star in the LMC. Early attempts to study \starname~by \citet{Moffat1989} lacked the sensitivity and time coverage necessary to uncover its multiplicity. Through multi-epoch spectroscopy in the near IR, \citet{Schnurr2008R136, Schnurr2009} showed that R~144 is a binary candidate, but could not establish a periodic behaviour. More recently, \citet{Sana2013} identified R~144 as a WNh+WNh double-lined spectroscopic binary (SB2) with a large RV amplitude (hence possibly high stellar masses), but the authors lacked the necessary time coverage to establish an orbital solution for this system. Given the total luminosity \mbox{$\log(L/L_\odot) \approx 6.8$} and large RV variations exhibited by the system, \citet{Sana2013} suggested that R~144 might be the most massive binary known in the Local Group.

This paper is the fifth paper in the framework of the Tarantula Massive Binary Monitoring (TMBM) project, a multi-epoch observing campaign design to constrain the orbital properties of about 100 massive binary candidates detected by the VLT-FLAMES Tarantula survey \citep[VFTS,][]{Evans2011, Sana2013b}. While \starname\ was not part of VFTS as it was used as a guiding star for the secondary guiding of the FLAMES plate in the VFTS, \starname\ was manually inserted in TMBM in view of its science interest.  

The first paper  of the TMBM series provided orbital solutions for 82 O-type systems in the sample \citep{Almeida2017}, the second focused on the very massive WR binary \object{R\,145} \citep{Shenar2017b}, while the third and fourth papers provided a spectroscopic and photometric analysis of double-lined spectroscopic binaries in the sample \citep{Mahy2020b, Mahy2020a}. 
In this study, we provide an orbital, spectroscopic, photometric, and polarimetric analysis of \starname. We do so relying on multi-epoch data acquired with the  FLAMES-UVES  and X-SHOOTER instruments mounted on the European Southern Observatory's (ESO) Very Large Telescope (VLT), described in Sect.\,\ref{sec:data}. Our analysis is presented in Sect.\,\ref{sec:analysis}, with the goal of 
deriving its orbit, disentangling its composite spectra, assessing the dynamical masses of both binary components, and deriving their physical parameters. In Sect.\,\ref{sec:discussion}, we discuss the implications of our results on our understanding of stellar evolution at the upper-mass end and of the upper-mass limit. We conclude our findings in Sect.\,\ref{sec:summary}.

\section{Data and reduction}\label{sec:data}

\subsection{Spectroscopic data}
\label{subsec:specdat}

\begin{figure*}
\centering
\includegraphics[width=\textwidth]{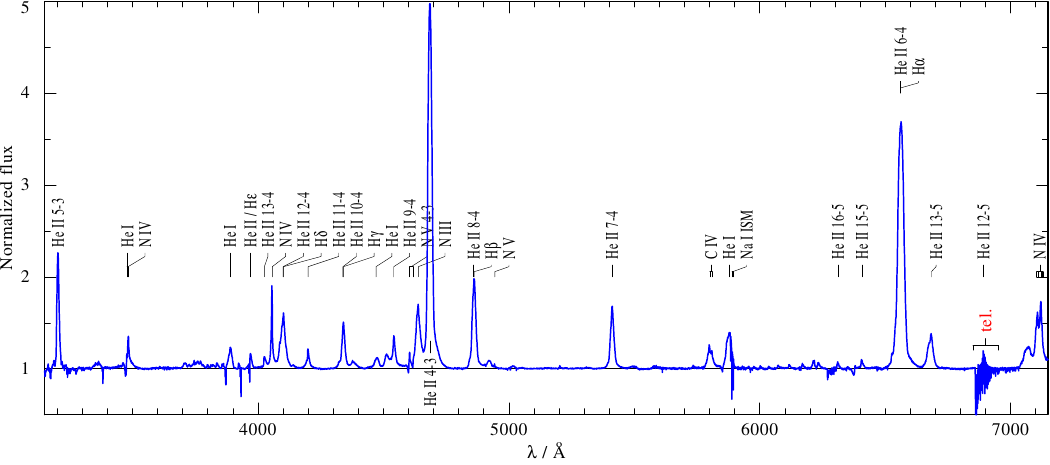}
\caption{X-SHOOTER spectrum of \starname~taken on 20 Feb 2014 (MJD = 56708.04), showing an overview of the optical part of the spectrum. } \label{fig:MegaSpec}
\end{figure*}

Our main spectroscopic dataset includes three different instruments spanning roughly 4\,yr. The first set of data was obtained with X-SHOOTER through the Dutch program for guaranteed observing time from Jan 2011 to Feb 2013, allowing the detection of the system as an SB2 binary and of large RV variations. The details of the data reduction and the early scientific results were presented in \citet{Sana2013}. \starname~was then monitored over a 1.5-yr period from Oct 2012 to Mar 2014 as the sole FLAMES-UVES target of the TMBM campaign, providing 27 useful observational epochs. The FLAMES-UVES data were obtained with the RED520 setup, providing a wavelength coverage from  4200 to 6800\,$\AA$ at a spectral resolving power of 47000. Each observation was composed of three consecutive 980s exposures for optimal cosmic removal. The data were reduced using the ESO FLAMES-UVES CPL pipeline under the \emph{esorex} environment. The sky subtraction was performed using the median value of three simultaneous sky spectra obtained through three individual FLAMES-UVES fibers located at empty locations in the 20'-diameter field-of-view of the FLAMES fiber positioner. 

A near real-time preliminary analysis of the FLAMES-UVES data allowed us to estimate the orbital period and time of periastron passage with sufficient accuracy to trigger a DDT monitoring of the 2014.2 periastron passage with the X-SHOOTER spectrograph. In this higher-cadence campaign, we obtained 13 X-SHOOTER spectra over a 10-d time base. The data were obtained in nodding-mode, with two nodding cycles completed to guarantee optimal sky subtraction. The nod-throw was 5" and a  jitter of 1" around each nodding position was included. NDITxDITs of 1x150s and 2x85sec were performed at each nodding position for, respectively, the UVB/VIS arms and the NIR arm. Narrow slit widths of 0.5" / 0.4" / 0.4" for the UVB / VIS / NIR spectra were adopted, delivering spectral resolving powers of 9700 / 18400 / 11600. The data were reduced following the same procedure as the X-SHOOTER GTO observations of \starname~\citep[see][]{Sana2013} and normalized through a polynomial fit to the continuum (see example in Fig.\,\ref{fig:MegaSpec}). The overall S/N after combining all nodding exposures of a given arm is in the range of 250 to 350. 

An additional spectrum obtained with The Fiber-fed Extended Range Optical Spectrograph (FEROS) instrument ($R = 48000, S/N \approx 60$, \citealt{Kaufer1997b}) mounted on the MPG/ESO-2.20m telescope in La Silla and reduced with the MIDAS FEROS pipeline. The data were acquired on 2011-05-16 in the framework of the Galactic O- and WN-star survey \citep[OWN,][]{Barba2010}, and were retrieved from the Library of Libraries of Massive-Star High-Resolution Spectra (LiLiMaRlin, \citealt{MaizApellaniz2019}). 
Finally, we retrieved two reduced UVES spectra from the ESO science portal. These were obtained in Nov 2003 with the DIC1~390+564 instrument setup and an entrance slit of 0.7", delivering a resolving power of $\sim$55\,000. The data provide continuous wavelength coverage from 3260 to 6680~\AA, but for a 40~\AA\ gap around 4560~\AA. The data are described in \citet{Crowther2011}.

In addition, we retrieved three spectra covering the far UV from the Mikulski Archive for Space Telescopes (MAST). 
% The orbital phases $\phi$ given in the following are based on our derived ephemeris (Sect.\,\ref{subsec:RVmeas}).  
One spectrum covers the spectral range $900-1200\,\AA$ and was obtained with the Far Ultraviolet Spectroscopic Explorer (FUSE), presented by \citet{Willis2004}.
The spectrum was acquired on 1999 Dec 16 (PI: Sembach, ID: P1031802000),
% (PI: Sembach, ID: P1031802000, $\phi = 0.16$),
and has a resolving power of $R\approx 20000$ and a signal-to-noise ratio of S/N $\approx 100$. We further retrieved two high-resolution spectra acquired with the International Ultraviolet Explorer (IUE). The first was acquired on 1978 Sep 29 (PI: Savage, ID: 	SWP02798), and
% $\phi=0.70$)   
covers the spectral range $1200-2000\,\AA$ with $R \approx 10000$ and S/N~$\approx 15$. The second was acquired on 1979 Feb 14 (PI:Savage, ID: LWR03766), 
% (PI:Savage, ID: LWR03766, $\phi = 0.57$), 
and covers $2000-3000\,\AA$ with $R \approx 10000$ and $S/N \approx 10$.  All three spectra are flux-calibrated, and are thus useful for fitting the spectral energy distribution (SED) of \starname, as well for deriving the wind parameters of both components.

%  orbital phase $\phi = 0.57$ with the ephemeris derived in Sect.\,\ref{sec:analysis})
% \EXPR IUE_SHORT = $OBSPATH // /UV/sp02798.dat
% \EXPR IUE_LONG = $OBSPATH // /UV/lr03766.dat

\subsection{Photometric data}
\label{subsec:photodat}
We obtained photometric data using the All-Sky Automated Survey for Supernovae (ASAS-SN) automated pipeline \citep{Shappee2014, Kochanek2017}. 
These data are available in the V, and more recently, $g$ filters. We chose to use the V dataset as it offers a broader bandwidth and the signal in this filter is more pronounced, covering a baseline of 1609\,d (roughly 22 orbital orbital cycles, see Sect.\,\ref{subsec:LC}).
% The g filter does not add much value to our attempts at light-curve modeling and has been ignored. 
We phase-folded the data on the orbital period derived in Sect.\,\ref{subsec:LC} and removed this signal from the light-curve. We then performed basic sigma clipping and removed clear alias/instrumental peaks which were present in the Fourier transform. This was done by a process, of phase-folding on the signal, templating this signal using a LOWESS filter \citep{Cleveland1979}, and removing this signal from the light-curve. After reductions were finished the initial orbital signal was added back to the light-curve. 

\starname~was also observed in several sectors of cycle~1 of the Transiting Exoplanet Survey Satellite (TESS) mission as part of the 30-min cadence full frame image (FFI) data \citep{Ricker2015}. More recently, \starname~was included as a target for 2-min cadence data as part of a TESS guest investigator proposal (PI: Bowman; GO3059). To date, 2-min cadence TESS data in sectors 27--30 are publicly available from the Mikulski Archive for Space Telescopes (MAST), which span a total of 108~d. We extracted a customised light-curve of \starname~from the 2-min TESS postage stamp pixel data using a 1-$\sigma$ threshold binary aperture mask, as defined in the \texttt{lightkurve} python package \citep{lightkurve2018}. The background flux estimates and barycentric timestamp corrections provided by the SPOC pipeline \citep{Jenkins2016} were also taken into account, and the light curve was normalised per sector by dividing through the median observed flux. To validate the extracted light-curve, we repeated the data reduction using the TESS FFI~data with different aperture masks, and visually compared the results.

In addition, we retrieved photometry using  VizieR\footnote{https://vizier.u-strasbg.fr/viz-bin/VizieR}: UBV photometry was retrieved from \citet{Parker1992}, R magnitude from \citet{Monet1998}, I magnitude from \citet{Pojmanski2002}, JHK and IRAC magnitudes from a compilation by \citet{Bonanos2009}, and WISE magnitudes from \citet{Cutri2014}.

\subsection{X-ray data}
\label{subsec:Xraydat}

X-ray observations of \starname~in the range $\approx$ 0.2 - 9\,keV  were acquired with the Chandra X-ray Observatory in the framework of the Chandra Visionary Program T-ReX (PI: Townsley). The observations were analysed with the ACIS Extract package \citep{Broos2010}. We refer the reader to \citet{Pollock2018} and \citet{Townsley2011} for a detailed account of the observing program and data reduction, respectively.

\subsection{Polarimetric data}
\label{subsec:poldat}

We collected linear polarimetric data for \starname~acquired and presented by \citet{Schnurr2009}. The data were obtained  between 1988 October and 1990 May at the 2.2-m telescope of the Complejo Astronomico El Leoncito (CASLEO) near San Juan, Argentina, and with Vatican Observatory Polarimeter \citep[VATPOL,][]{Magalhaes1984}.

\section{Analysis}\label{sec:analysis}

\subsection{Radial-velocity measurements and orbital solution}\label{subsec:RVmeas}

\begin{figure}
\centering
\includegraphics[width=.5\textwidth]{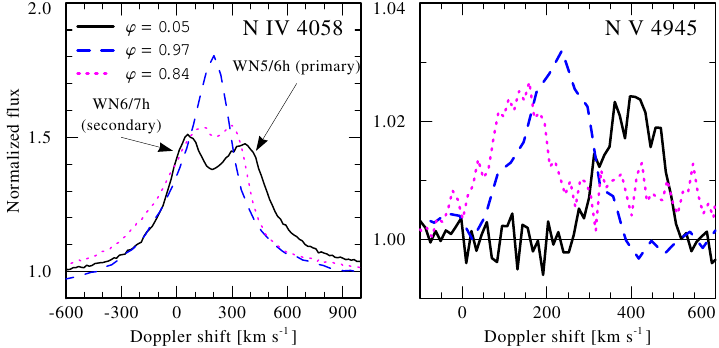}
\caption{Plotted are three X-SHOOTER spectra (MJD = 55452.89, 55585.65, 56708.04, see legend) corresponding to phases close to the radial velocity extremes ($\phi = 0.05, 0.84$ with ephemeris derived in this study) and the systemic velocity ($\phi = 0.96$), and focusing on the \NIV~and \NVred~lines. Both binary components are seen in the \NIV~line, but only the hotter primary is seen in the \NVred~line. Conversion to Doppler space is with reference to rest wavelengths.} 
\label{fig:MegaSpecZoom}
\end{figure}

\begin{figure}
\centering
\includegraphics[width=.5\textwidth]{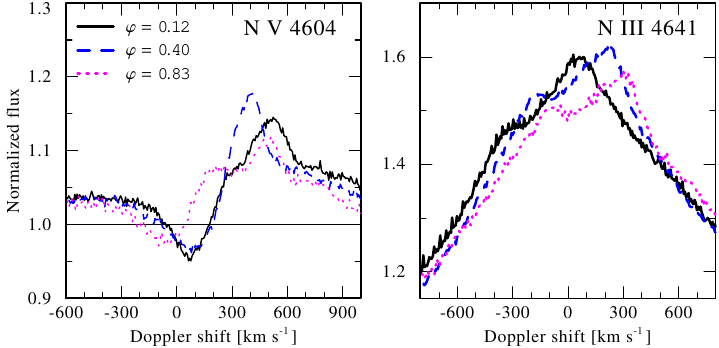}
\caption{As Fig.\,\ref{fig:MegaSpecZoom}, but showing three FLAMES-UVES spectra (MJD = 56645.57, 56645.57, 56697.69), and focusing on the \NVblue~and \NIII~lines. Both components contribute to the \NVblue~lines, while the cooler secondary dominates the \NIII~lines.
}
\label{fig:MegaSpecZoom2}
\end{figure}

\begin{figure*}
\centering
\begin{tabular}{ccc}
\includegraphics[width=0.31\textwidth]{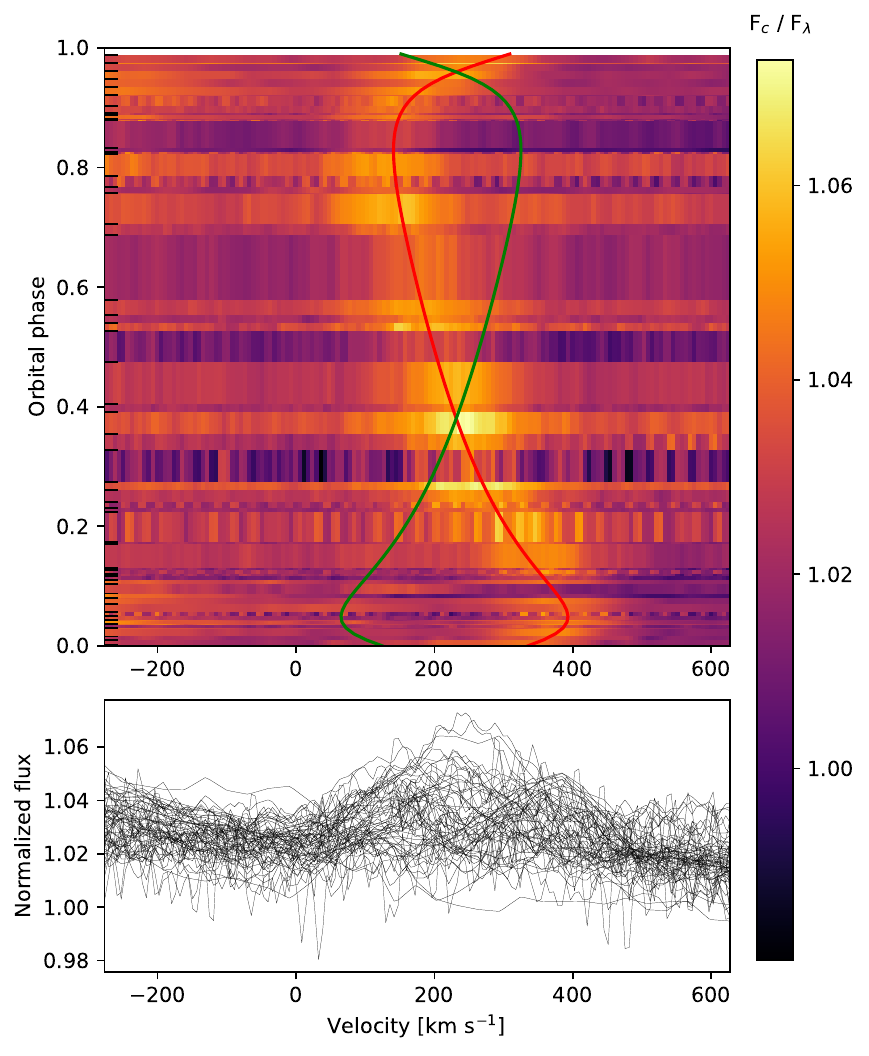} & 
\includegraphics[width=0.3\textwidth]{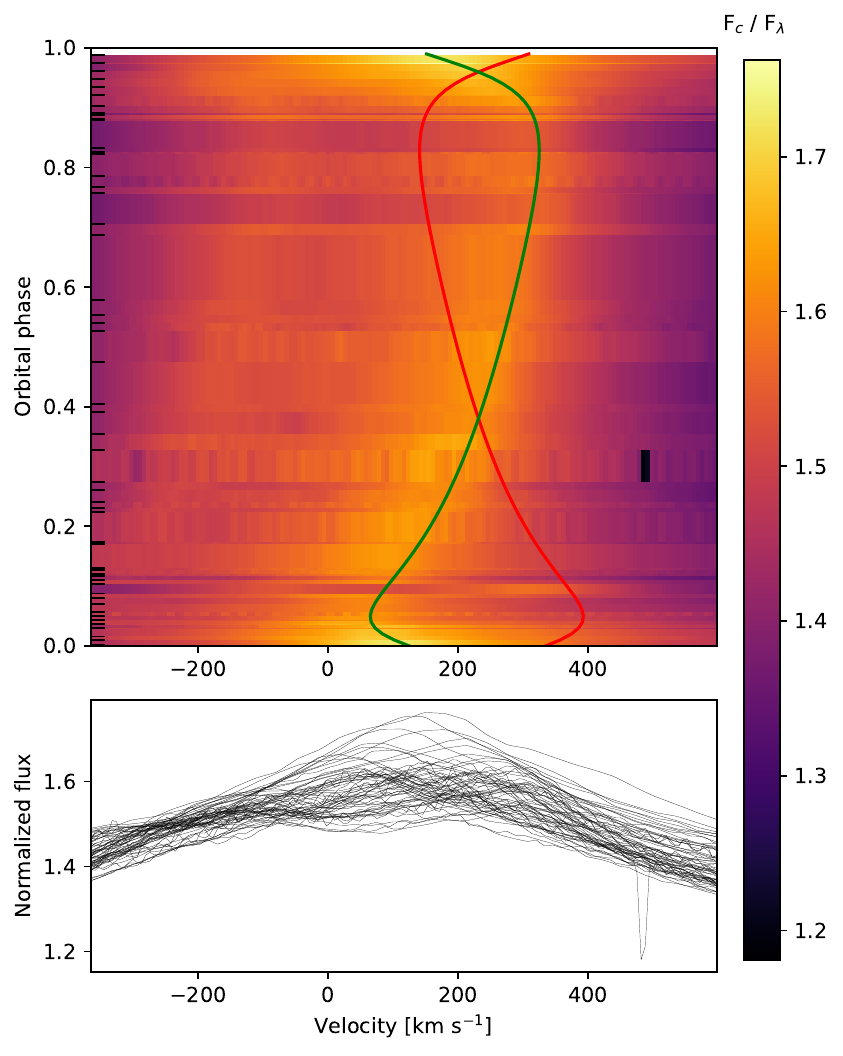} & 
\includegraphics[width=0.31\textwidth]{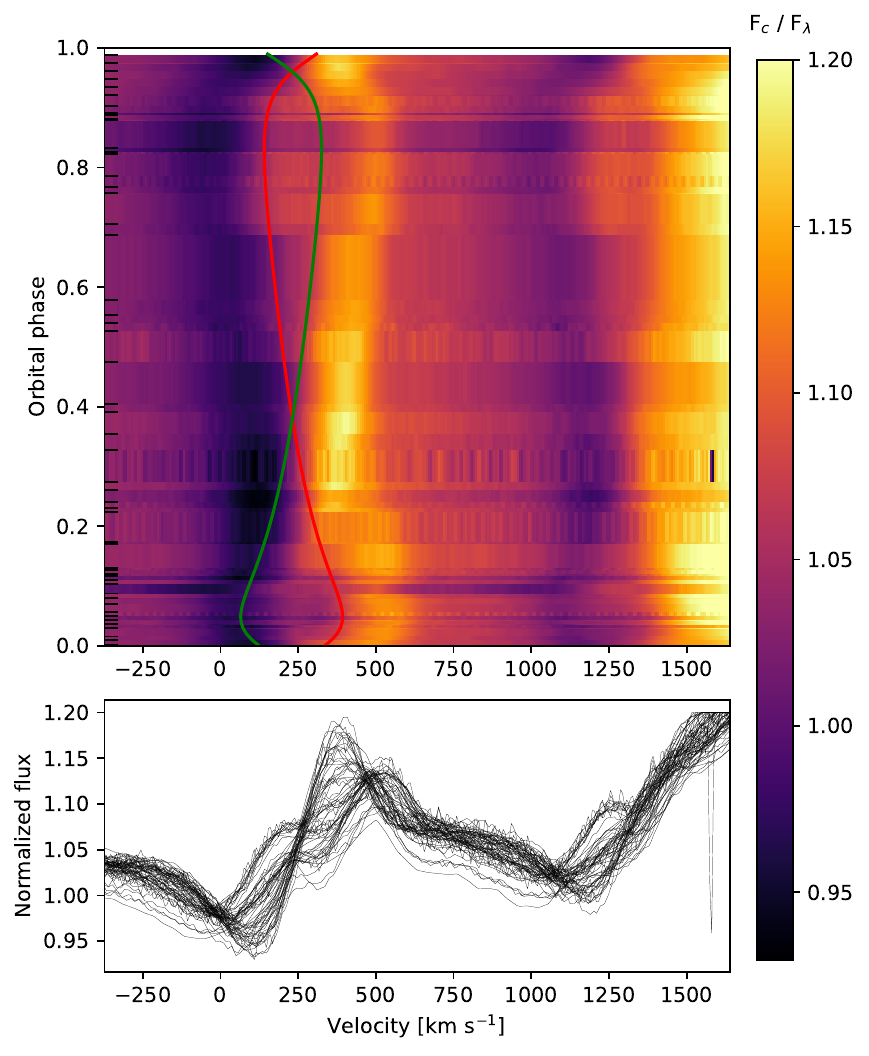} 
\end{tabular}
\caption{Dynamical spectra for the \NVred, \NIII, and \NVblue~lines. Red and green curves depict the orbital solution derived in Sect.\,\ref{subsec:RVmeas} for the primary and secondary components, respectively.
}
\label{fig:dynspec}
\end{figure*}

The optical spectrum of \starname~is dominated by emission lines belonging primarily to He and N, as is typical for WN stars (Fig.\,\ref{fig:MegaSpec}).
A closer inspection of the N\,{\sc iv}\,$\lambda 4058$ and N\,{\sc v}\,$\lambda 4945$ lines reveals the presence of two WR stars in the system (Fig.\,\ref{fig:MegaSpecZoom}). The two stars exhibit similar spectra, although one star appears to be slightly hotter. This is evident when examining, for example, the N\,{\sc v}\,$\lambda 4945$ line, which forms predominantly in one of the components. Another example can be seen in Fig.\,\ref{fig:MegaSpecZoom2} for the  N\,{\sc v} $\lambda \lambda 4604,4620$ doublet and the \NIII~triplet. Both stars appear to contribute in both line blends, but the indicative anti-phase behaviour of these lines implies that each of them is dominated by a different stellar component. The dynamical spectra of the \NVred, \NIII, and \NVblue~lines point towards similar conclusions (Fig.\,\ref{fig:dynspec}). Throughout this paper, 
we refer to the hotter component in \starname~as the primary star.

The RVs are measured via cross-correlation \citep{Zucker1994}. The formalism is outlined in \citet{Shenar2019} and \citet{Dsilva2020}. As a first step, a single observation is used as a template against which the remaining observations are cross-correlated to obtain the relative RVs. In the second step, a new template is formed by co-adding all observations in the same frame. This template is then used to re-measure the RVs, and the process is repeated until convergence is attained (typically 2-3 iterations). 

For the (hotter) primary, we use the clearly isolated \NVred~line (see Fig.\,\ref{fig:MegaSpecZoom}). For the (cooler) secondary, we use the peak of the \NIII~triplet (see Fig.\,\ref{fig:MegaSpecZoom2}), as it is strongly dominated by the secondary. The template of the primary, focused on the \NVred~line, is calibrated to the rest-frame by comparing it with our model atmosphere (Sect.\,\ref{subsec:nonLTE}), hence enabling us to convert relative RVs to absolute values for the primary. As the positions of the line centroids depend on the wind parameters, providing absolute RVs for WR stars is model-dependent, and we estimate an accuracy of $\approx 20$\,\kms~on the absolute value from comparison to models. The final measurements are given in Table\,\ref{tab:log}.

Relying on the RV measurements, we derive an orbital solution, constraining the  time of periastron $T_0$, eccentricity $e$, argument of periastron $\omega$, systemic velocity $V_0$, and RV amplitudes $K_1$ and $K_2$. The orbital period is fixed to $P=74.2074$-d based on our light-curve analysis, which is described in Sect.\,\ref{subsec:LC}. The remaining orbital parameters are fitting parameters. For the minimization, we use a self-written Python script that relies on the minimization package lmfit\footnote{https://lmfit.github.io/lmfit-py} \citep{Shenar2019}, applying the differential evolution algorithm  \citep{Newville2014}.
Since the RVs of the secondary are relative, we allow for an additional constant offset between the two RV sets to allow an orbital solution with a single systemic velocity. We fit the RV curves simultaneously with our light-curve model, which is described in Sect.\,\ref{subsec:LC}. However, no notable differences are obtained when minimizing both separately.

\begin{figure}
\centering
\includegraphics[width=.5\textwidth]{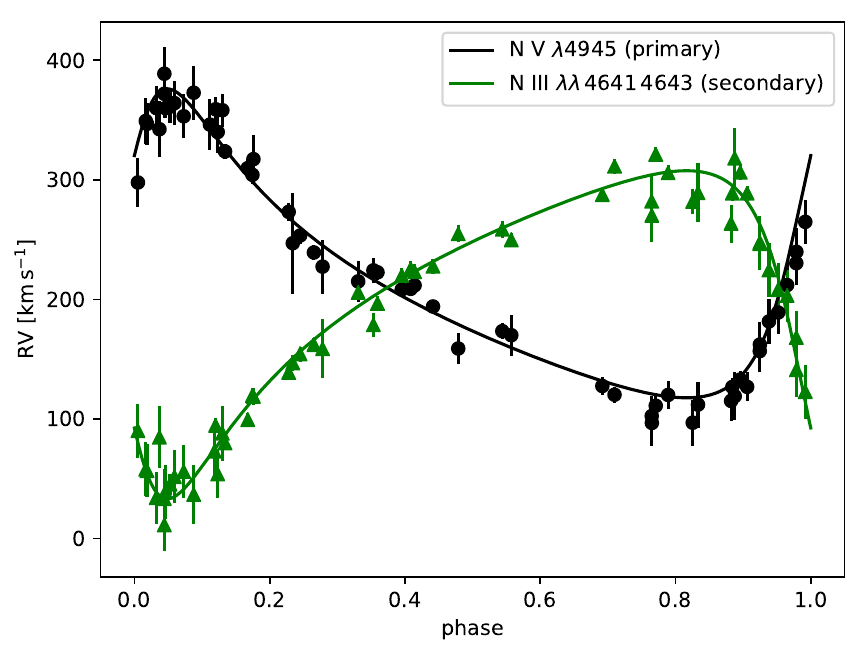}
\caption{RV measurements of the \NVred~line (primary) and the \NIII~triplet (secondary), along with the RV curves corresponding to the orbital parameters given in Table\,\ref{tab:parameters}.} 
\label{fig:orbit}
\end{figure}

The derived orbital parameters are given in Table\,\ref{tab:parameters}, and the best-fitting orbital solution is shown in Fig.\,\ref{fig:orbit}. We obtain a root mean square error (rms) of 15\,\kms~for both components. In contrast, the formal measurement errors have a mean of $\sigma \approx 10\,$\kms. It is, however, not uncommon for formal errors of RV measurements to be underestimated in the case of WR stars, primarily due to spectral line variability that stems from either intrinsic wind variability or wind-wind collisions in the system. As the Keplerian model should provide a virtually perfect description to time variability of the RVs, we multiply the formal errors by the ratio of the measured rms value and the mean of the formal errors, since these scaled errors should reflect more realistically the true measurement errors.

The inferred orbital parameters are given in Table\,\ref{tab:parameters}. The two stars are found to have almost identical RV-amplitudes and hence masses, with an indication for the hotter companion being slightly more massive. The system is eccentric and seems to strongly resemble the properties of other known WNh+WNh binaries in the LMC such as \object{R\,145} \citep{Shenar2017b} and \object{Mk\,34} \citep{Tehrani2019}. The minimum masses of the primary and secondary are $M \sin^3 i = 48.3\,M_\odot$ and  $45.5\,M_\odot$, respectively.

    % K1:      125.916981 +/- 2.16882582 (1.72%) (init = 117)
    % K2:      129.512680 +/- 2.82220207 (2.18%) (init = 138)

\subsection{Spectral disentangling}\label{subsec:disen}

Spectral disentangling is a mathematical technique that separates a series of phase-dependent spectra of a multiple system to the individual spectra of the stellar constituents \citep[e.g.,][]{Hadrava1995}. The appearance of the separated spectra depends on the orbital parameters (or individual RVs of the components) as an input. Here, we use the shift-and-add technique to separate the spectra \citep{Marchenko1998, Gonzalez2006}. The technique has been used in the past for the study of WR binaries \cite[e.g.,][]{Demers2002, David-Uraz2012, Shenar2017b, Shenar2018, Shenar2019}. 
As all orbital parameters were derived in Sect.\,\ref{subsec:RVmeas}, we can in principle fix them to separate the spectra. However, by allowing $K_1$ and $K_2$ to vary, we can use additional line blends to verify the results obtained in Sect.\,\ref{subsec:RVmeas}.  We do this by calculating the $\chi^2$ of the residuals obtained when shifting-and-adding the disentangled spectra and subtracting the resulting spectrum from the individual observations iteratively \citep[see e.g.,][]{Shenar2020LB1}. 

\begin{figure}
\centering
\begin{tabular}{c}
\includegraphics[width=0.465\textwidth]{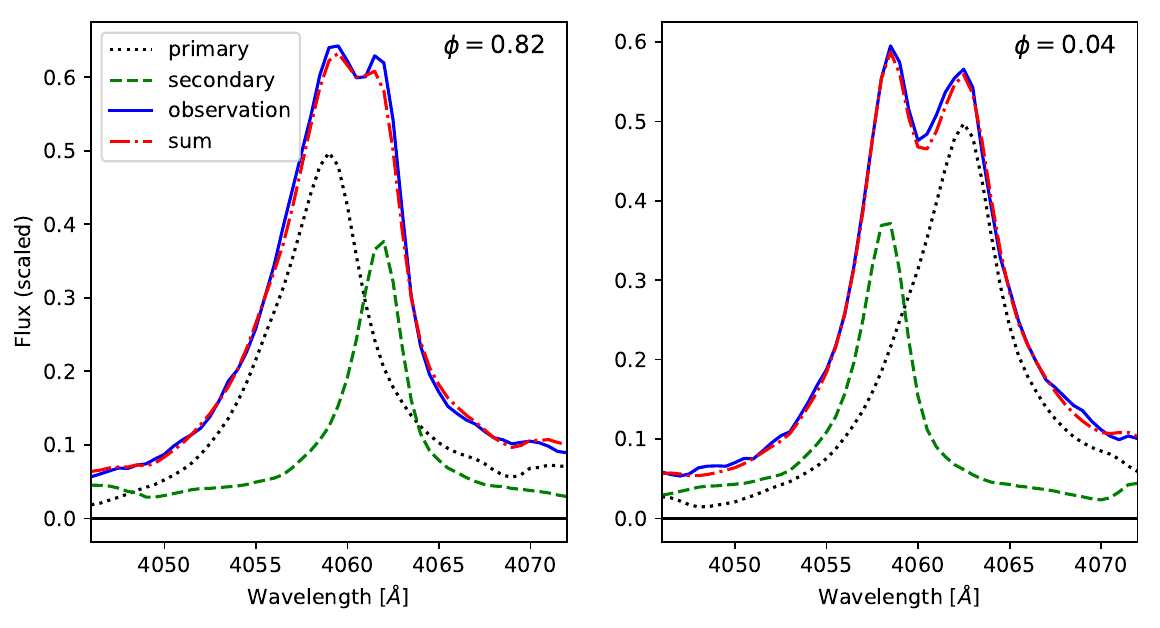}\\
\includegraphics[width=0.47\textwidth]{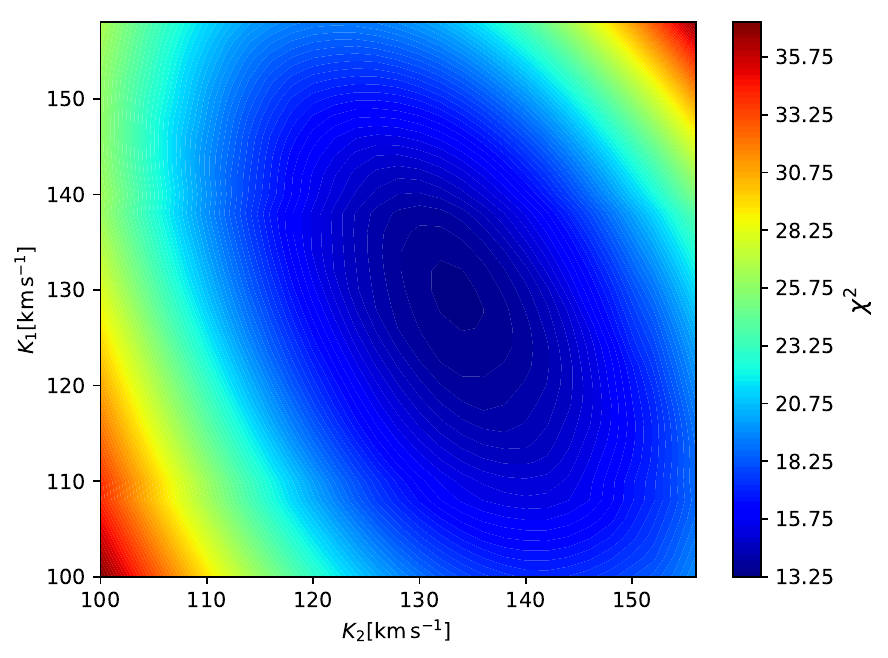}
\end{tabular}
\caption{{\it Upper panels:} comparison between the disentangled spectra obtained for the best-fitting $K_1$ and $K_2$ values, their sum, and observations at two phases close to RV extremes for the N\,{\sc iv}\,$\lambda 4058$. {\it Lower panel:} reduced $\chi^2$ as a function of $K_1$ and $K_2$ for the N\,{\sc iv}\,$\lambda 4058$ line. The analysis is performed using the X-SHOOTER data. The minimum is at $K_1 = 129 \pm 6\,$\kms, $K_2 = 134\pm 4$\,\kms.}
\label{fig:NIVdis}
\end{figure}

\begin{figure}
\centering
\begin{tabular}{cc}
\includegraphics[width=0.465\textwidth]{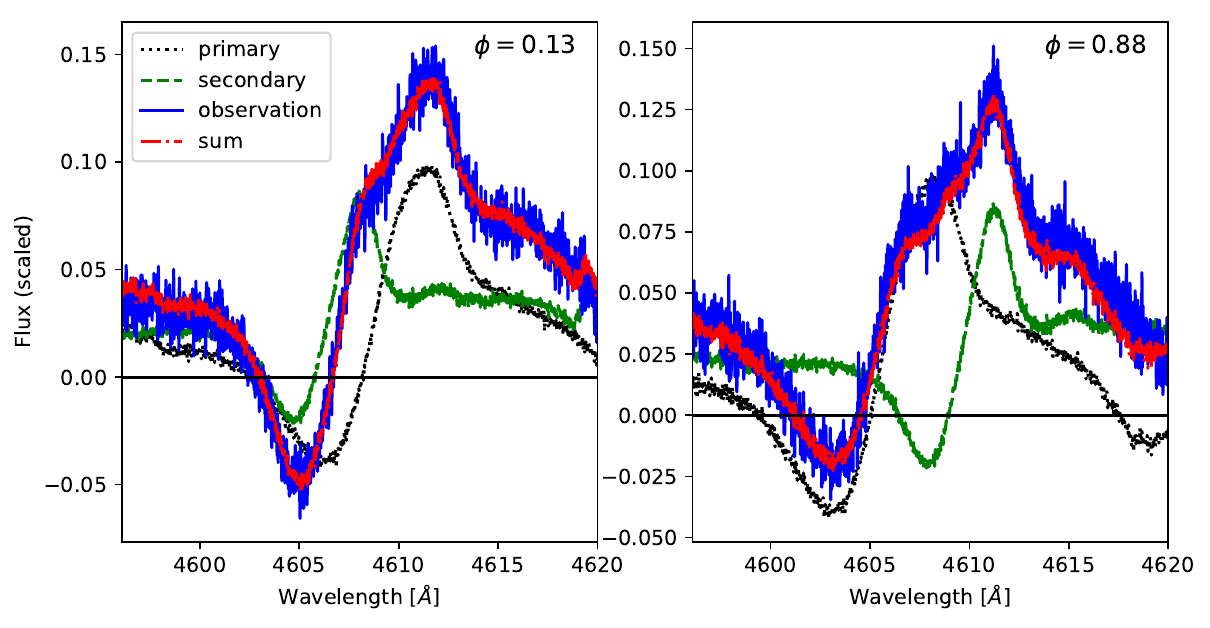}\\
\includegraphics[width=0.47\textwidth]{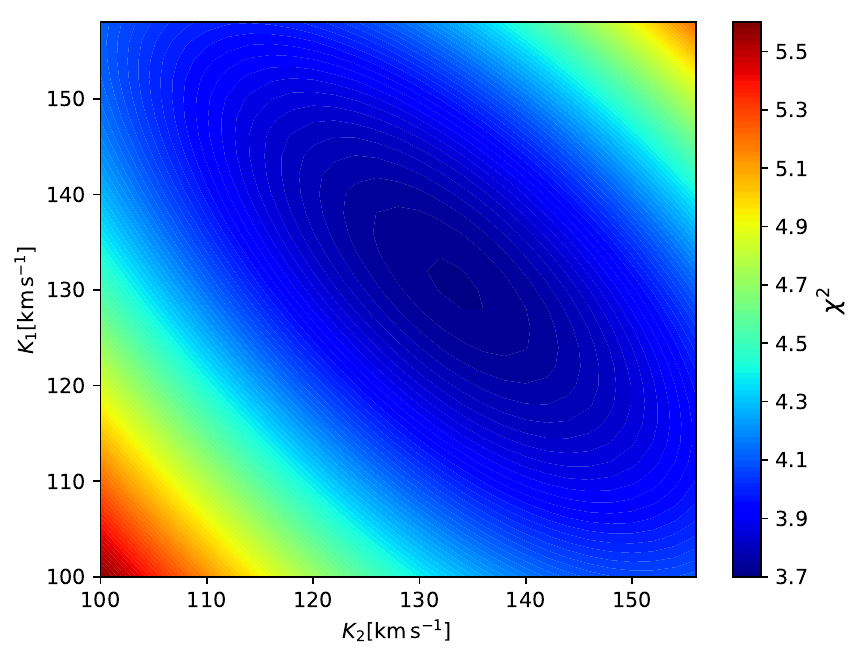}
\end{tabular}
\caption{As Fig.\,\ref{fig:NIVdis}, but for \NVblue. The analysis is performed using the FLAMES-UVES data and X-SHOOTER data.  The minimum is at $K_1 = 130 \pm 3\,$\kms, $K_2 = 134\pm 4$\,\kms. 
}
\label{fig:NVdis}
\end{figure}

In Fig.\,\ref{fig:NIVdis}, we show for the  N\,{\sc iv}\,$\lambda 4058$ line the match between two observations close to RV extremes and the sum of the disentangled spectra shifted according to the best-fitting $K_1$ and $K_2$ values. The analysis relies on the X-SHOOTER data, which cover the relevant wavelength regime. The lower panel shows the reduced $\chi^2$ map, where a clear minimum consistent with the values given in Table\,\ref{tab:parameters} is shown.  Figure\,\ref{fig:NVdis} shows the same as Fig.\,\ref{fig:NIVdis}, but for the \NVblue~lines and using both the X-SHOOTER and FLAMES-UVES spectra. Again, the results agree very well (within 1$\sigma$) with our measurements from the RV curves.  
Overall, our three measurements of $K_1$ and $K_2$ are consistent within their respective 1$\sigma$ error measurements and indicate that the hotter primary is slightly more massive than its cooler companion.

\begin{figure}
\centering
\includegraphics[width=.5\textwidth]{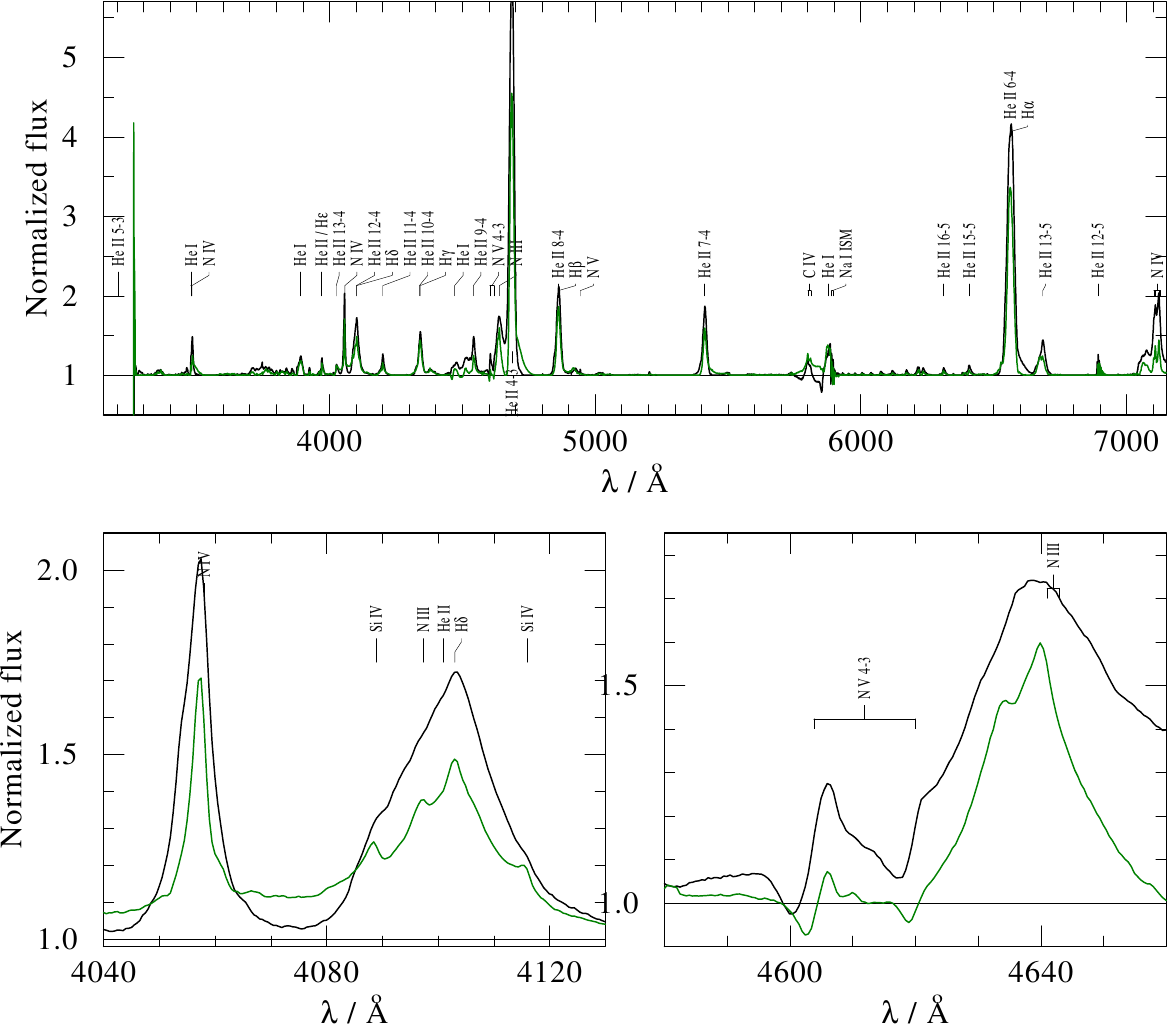}
\caption{Disentangled spectra of \starname~obtained with the shift-and-add technique for the hotter primary (WN5/6h, black line) and cooler secondary (WN6/7h, green line).
}
\label{fig:disspec}
\end{figure}

The disentangled X-SHOOTER spectra are shown in Fig.\,\ref{fig:disspec}. The spectra can be disentangled up to a scaling factor which depends on the optical light ratio of the two stars. For reasons that will be outlined in Sect.\,\ref{subsec:nonLTE}, we adopt the optical light ratio of $f_1/f_2(V) = 0.79$. The spectra shown in Fig.\,\ref{fig:disspec} are scaled accordingly. The process of disentangling, especially of two emission-line stars and in the presence of non-Keplerian variability (e.g., wind-wind collisions), can introduce spurious features that should not be over-interpreted. For example, the shape of the electron-scattering wings (notably H and He\,{\sc ii} lines) and P-Cygni lines (notably He\,{\sc i} lines), as well as the absolute strengths of the lines, can be quite degenerate in the solution. Hence, we primarily rely on the disentangled spectra to classify the stars and compare them qualitatively.

While the two spectra are similar, a few differences are apparent. Both components show all three ionisation stages N\,{\sc iii, iv, v}, but the hotter primary shows N\,{\sc v} more prominently, while the cooler secondary dominates in N\,{\sc iii} lines, as anticipated. The spectral features of the cooler secondary tend to be sharper, whereas those of the primary more smeared, pointing towards slightly larger terminal velocity $\varv_\infty$  for the hotter primary.

Following the quantitative classification schemes by \citet{Smith1996}, the primary falls in-between the WN5h and WN6h classes, while the secondary falls in-between the WN6h and WN7h classes. This is also confirmed by morphological comparisons with spectra presented by \citet{Crowther1997} and \citet{Crowther1998}. We therefore classify the primary as WN5/6h and the secondary as WN6/7h.  The '-h' suffix stems from the ratio of Balmer to He\,{\sc ii} lines, which suggests that both still have a significant amount of hydrogen in their envelopes. 

\subsection{Spectral analysis} \label{subsec:nonLTE}

\begin{table}
\centering
    \caption{Inferred parameters for \starname~from the orbital/light-curve/polarimetric (upper part, Sects.\,\ref{subsec:RVmeas}, \ref{subsec:LC}, \ref{subsec:Polan}.) spectral (middle part, Sect.\,\ref{subsec:nonLTE}), and evolutionary (lower part, Sect.\ref{subsec:evolution}) analyses. }
    \begin{tabular}{lcc} \hline \hline
        Parameter & primary & secondary \\ 
        Spectral type    & WN5/6h & WN6/7h \\           
        \hline
        \multicolumn{3}{c}{orbital analysis}   \\       
        \hline      %   74.20735785953177
         $P_\mathrm{orb}$\,[d]\,$^{\rm a}$ & \multicolumn{2}{c}{$74.2074 \pm 0.0043$} \\
         $T_0$\,[MJD]  & \multicolumn{2}{c}{$58268.98 \pm 0.07$} \\
         $e$  & \multicolumn{2}{c}{$0.506\pm 0.004$} \\
         $V_0$\,[\kms{}]\,$^{\rm b}$ &  \multicolumn{2}{c}{210$\pm20$ } \\         
        %  $\omega$  & \multicolumn{2}{c}{$302.2 \pm 2.0$} \\   
         $\omega$  & $304.6 \pm 0.8$ & $124.6 \pm 0.8$ \\       
         $\Omega$\,$^{\rm c}$  & \multicolumn{2}{c}{$114 \pm 9$} \\
         $K$\,[\kms{}]  &  $129.5 \pm  2.3$ & $137.4 \pm 2.1$ \\
         $M\,\sin^3 i$\,[\msun] &  $48.3 \pm 1.8$ & $45.5 \pm 1.9$ \\
         $a \sin i$\,[$R_\odot$]&  $164.0 \pm 3.0$ & $174.3 \pm 2.7$ \\              
         $q(\frac{M_2}{M_1})$  &   \multicolumn{2}{c}{$0.94 \pm 0.02$} \\    
         $i$\,[deg]\,$^{\rm d}$ &  \multicolumn{2}{c}{60.4$ \pm $1.5}\\            
         $a$\,[$R_\odot$]&  $189 \pm 5$ & $200 \pm 4$ \\               
         $M_{\rm dyn}$\,[$M_\odot$] &  $74 \pm 4$ & $69 \pm 4$ \\   
        %  \hline
        % \multicolumn{3}{c}{orbital analysis (disentangling of \NVblue)}  \\       
        % \hline                 
        %  $K$\,[\kms{}] &  $136 \pm  4$ & $143 \pm 4$ \\
        %  $M\,\sin^3 i$\,[\msun] &  $55.0 \pm 3.5$ & $52.3 \pm 3.4$ \\     
        %  $q(\frac{M_2}{M_1})$ &   \multicolumn{2}{c}{$0.95 \pm 0.04$} \\      
        % %  $a \sin i$\,[R$_\odot$] &  $171 \pm 4$ & $179 \pm 6$ \\    
        %  $a$\,[$R_\odot$] &  $199 \pm 7$ & $210 \pm 7$ \\              
        %  $M_{\rm dyn}$\,[$M_\odot$] &  $85 \pm 7$ & $81 \pm 7$ \\ 
        \hline
        \multicolumn{3}{c}{spectral analysis}   \\       
        \hline        
        $f/f_{\rm tot}(V)$\,$^{\rm e}$ & 0.44 (fixed)  & 0.56 (fixed)  \\    
        $T_*$\,[kK] & $50\pm2$ & $45\pm2$   \\
        $T_{2/3}$\,[kK] & $45\pm2$ & $40\pm2$   \\        
        $\log R_{\rm t}$\,[R$_\odot$] & $1.10\pm0.05$ & $1.15\pm0.05$   \\ 
        $\log\,L$\,[L$_\odot$]  & $6.44\pm0.05$  & $6.39\pm0.05$  \\  % 3.2
        $M_{V, {\rm Smith}}$ [mag] & $-7.3\pm0.2$  & $-7.6\pm0.2$  \\            
        $E_{B-V}$\,[mag]  & \multicolumn{2}{c}{$0.20\pm0.01$}   \\      
        $R_*$\,[R$_\odot$] & $22 \pm 3$ & $26\pm3$  \\ % 5.9 
        $R_{2/3}$\,[R$_\odot$] & $27 \pm 3$ & $31\pm3$  \\ % 5.9 
        $D$ & $10$ (fixed) & $10$ (fixed)  \\ % 5.9 
        $\xi(R_*)$ & $30$ (fixed) & $30$ (fixed)  \\ % 5.9         
        $\log \dot{M}$\,[\smy] & $-4.38 \pm 0.15$ & $-4.34 \pm 0.15$  \\ % 5.9 
        $\varv_\infty$\,[\kms{}] & $1400 \pm 100$ & $1200 \pm 100$  \\ % 5.9
        $X_{\rm H}$ & $0.35\pm0.05$ & $0.40\pm0.05$  \\ % 5.9 
        $X_{\rm C} / 10^{-4}$ & $7\pm3$ & $7\pm3$   \\ % 5.9 
        $X_{\rm N} / 10^{-4}$ & $40_{-10}^{+20}$ & $40_{-10}^{+20}$  \\ % 5.9 
        $X_{\rm O} / 10^{-4}$ & $\lesssim 0.1$ & $\lesssim 0.1$  \\ % 5.9 
        $\varv \sin i$\,[\kms{}] & $\lesssim 80$ & $\lesssim 100$  \\
        $\varv_{\rm eq}$\,[\kms{}]\,$^{\rm f}$ & $\lesssim 100$ & $\lesssim 120$ \\ 
\hline        
        \multicolumn{3}{c}{evolution models\,$^{\rm g}$}   \\   
\hline       

        % $M_{\rm ev, ini}$\,[\msun] &  $130\pm13$ & $119\pm12$ \\ 
        $M_{\rm ev, ini}$\,[\msun] &  $148\pm14$ & $132\pm20$ \\         
        $\varv_{\rm eq, ini}$\,[\kms] &  $90^{+315}_{-62}$ & $10^{+36}_{-15}$ \\  
        $M_{\rm ev, cur}$\,[\msun] &  $111\pm12$ & $100\pm12$ \\       
        $\varv_{\rm eq, cur}$\,[\kms] &  $\le 20$ & $\le 10$ \\            
        age [Myr] &  $1.96\pm0.21$ & $2.10\pm0.20$ \\  
    \hline
    \end{tabular}
\tablefoot{
$^{\rm a}$ Obtained with the ASAS-SN and TESS data using the \mbox{pwkit pdm} algorithm (Sect.\,\ref{subsec:LC}). $^{\rm b}$ The absolute calibration of the systematic velocity is model-dependent and is  obtained through calibration with model spectra. The 20\,\kms~uncertainty is a rough estimate.   $^{\rm c}$ Obtained from polarimetry (Sect.\,\ref{subsec:Polan}). $^{\rm d}$ Obtained from light-curve analysis (Sect.\,\ref{subsec:LC}). $^{\rm e}$ Fixed based on mass-luminosity calibrations for homogeneous stars relying on the derived mass ratio. $^{\rm f}$ Assuming the orbital and rotational axes are aligned. $^{\rm g}$ Inferred with BONNSAI using tracks calculated by \citet{Graefener2021}. 
 }    
\label{tab:parameters}
\end{table}

\begin{figure*}[!htp]
\centering
\includegraphics[width=0.85\textwidth]{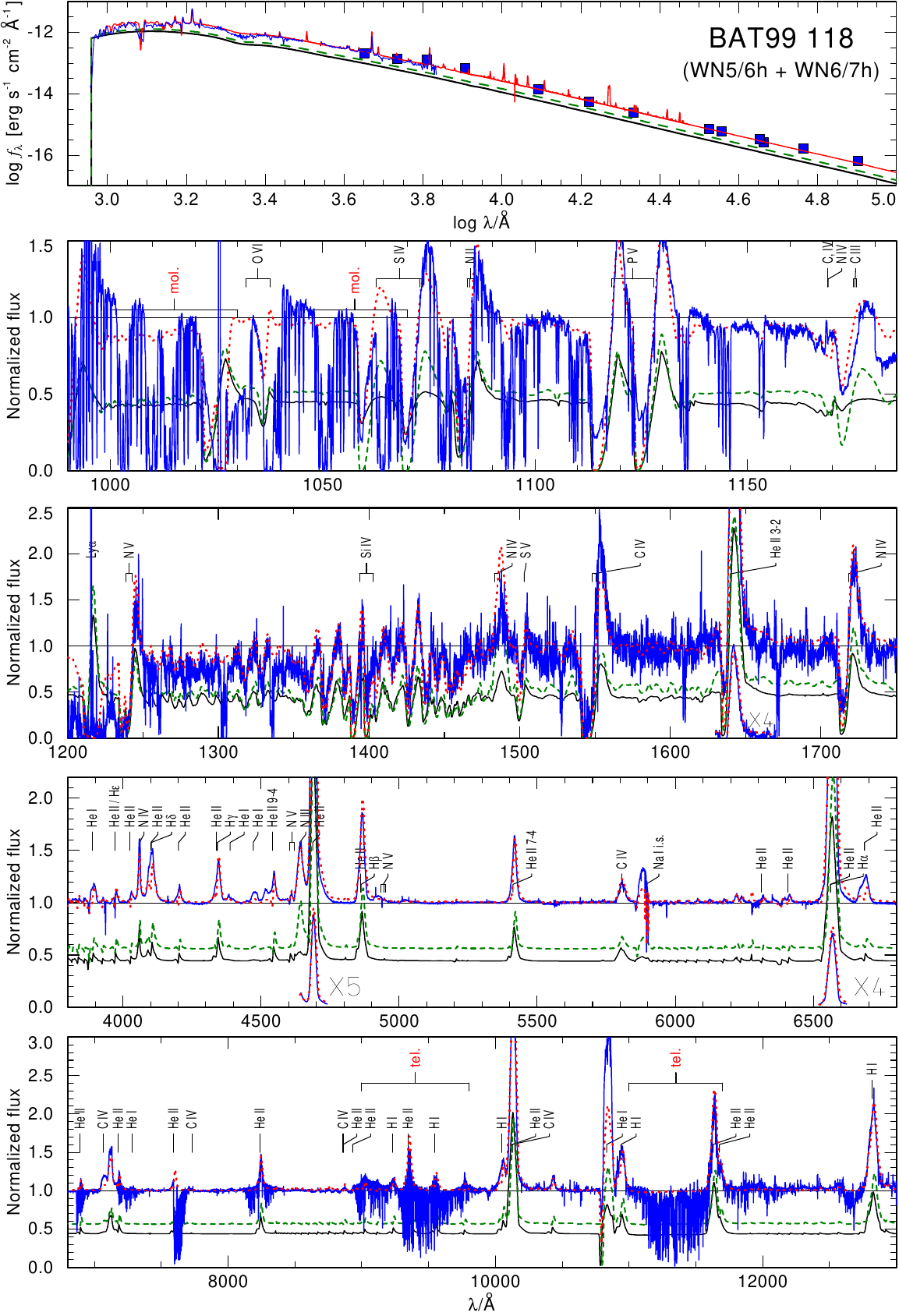}
\caption{Comparison between observed SED (blue squares and lines, upper panel) and normalized FUSE, IUE, and X-SHOOTER ($\phi=0.76$) spectra (lower panel)
and the synthetic composite spectrum (red dotted line). The composite spectrum is the sum of the hotter primary (black solid line) and cooler  secondary (green dashed line). Prominent telluric and molecular bands are marked in red.
% The observed and modeled spectra in the UV are binned at 1 Å for clarity. Lines that are strongly affected by WWC are marked with
% red idents.
}
\label{fig:specan}
\end{figure*}

% \newpage

To derive the physical parameters of both stellar components, we perform a spectroscopic analysis of the system while relaxing the assumption of local thermodynamic equilibrium (non-LTE). We use the Potsdam Wolf-Rayet (PoWR) model atmosphere code \citep{Hamann2003, Graefener2002, Sander2015}. PoWR is a 1D code that solves the radiative transfer problem in spherical geometry. It was originally developed for the analysis of WR stars, but is  applicable to any hot star with an expanding atmosphere \citep[e.g.,][]{Shenar2015, GimenezGarcia2016, Ramachandran2019}. While PoWR is designed for the analysis of single stars, the analysis of binaries is possible by relying on the disentangled spectra (Sect.\,\ref{subsec:disen}) and on the sum of individual models calculated for the two binary components. Even though this neglects the impact of wind-wind collisions (WWC), mutual irradiation, or other non-spherical effects, these effects are globally negligible, although they can dominate in the case of individual lines (e.g., the He\,{\sc i} lines, see Sect.\,\ref{subsec:WWC}). 

We calculate tailored models for our analysis, but we strongly rely on PoWR grids\footnote{http://www.astro.physik.uni-potsdam.de/PoWR} calculated for LMC metallicities \citep{Hamann2004, Todt2015} for error estimations. We include model atoms for H, He, C, N, O, P, Si, S, and the iron group elements (dominated by Fe). The mass fractions of P, Si, S, and Fe are fixed similarly to \citet{Hainich2014} and \citet{Shenar2019}: 
$X_{\rm P} = 2.91 \times 10^{-6}$, 
$X_{\rm Si} = 3.21 \times 10^{-4}$, 
$X_{\rm S} = 1.55  \times 10^{-4}$,
$X_{\rm Fe} = 7.02 \times 10^{-4}$; the rest are fitting parameters. The goodness of the fit is judged through visual inspection; producing $\chi^2$-like estimates in the framework of full non-LTE calculations, especially in a binary, is not feasible and would anyhow neglect systematic errors originating in the uncertain structure of the winds, which are the dominant factor of uncertainty here.

The main parameters defining a model atmosphere are the effective temperature $T_*$, the luminosity $L$, the mass-loss rate $\dot{M}$, and the chemical abundances, expressed here as mass fractions. Because of the optically-thick winds of WR stars, their photospheres  (defined at a mean optical depth of $\tau_{\rm ross} \approx 2/3$) are typically located well above their hydrostatic stellar surfaces. The inner boundary of the model, referred to as the stellar
radius $R_*$, is defined at $\tau_{\rm Ross} = 20$.\footnote{$\tau = 20$ is a standard convention; any value above $\approx 10$ would be appropriate due to the exponential increase of $\tau(r)$} The stellar radius $R_*$ is related to $T_*$ and $L$ via the Stefan-Boltzmann equation $L/L_\odot = (R/R_\odot)^2\,(T/T_\odot)^4$. The wind velocity $\varv(r)$ assumes the functional form of a $\beta$-law \citep{Castor1975}, characterised by a unitless parameter $\beta$ of the order of unity and the terminal velocity $\varv_\infty$. We explored various $\beta$ values in the range 1-10, and can conclude that $\beta \lesssim 3$, which is consistent with previous literature for late-type WN stars \citep[e.g.,][]{Chene2008}. We therefore adopt the standard value of $\beta =1$, lacking evidence to assume otherwise. The Doppler widths of the opacity and emissivity profiles are defined as $\varv_{\rm Dop} = \sqrt{\varv_{\rm th}^2 + \xi^2}$, where $\varv_{\rm th}(r)$ is the thermal motion of the chemical species, and $\xi(r)$ is a prespecified microturbulence \citep{Shenar2015}. Here, we assume $\xi(R_*) = 30\,$\kms~and that $\xi(r)$ grows with the wind velocity with a proportionality factor of 10\%. 

Winds of massive stars are not homogeneous but are rather clumped 
\citep{Moffat1988, Hillier1991, Lepine1999, Puls2006, Oskinova2007, Sundqvist2010}.
Optically thin clumping (microclumping) is accounted for here by introducing the clumping factor $D$, which describes the density factor of clumped material compared to the equivalent smooth wind ($D = 1/f$, where $f$ is the filling factor). In ideal cases, $D$ can be derived by simultaneously considering spectral features whose strength is proportional to the density $\rho$ (P-Cygni lines, electron-scattering wings) and $\rho^2$ (recombination lines, free-free emission). However, given the uncertainties in the disentangling procedure (see Sect.\,\ref{subsec:disen}) and the ambiguity in associating UV features with the individual binary components, this is not feasible here. While we explored various clumping parameters up to $D=100$, we cannot derive $D$ unambiguously, and hence fix the clumping factor to the typical value of $D=10$ \citep[e.g.,][]{Hainich2014, Shenar2019}. The mass-loss rates derived can be easily scaled to other clumping factors by preserving the product $\dot{M} \sqrt{D}$.

For a given value of $T_*$ and chemical abundances, the strength of emission lines in visual wavelengths depends not only on $\dot{M}$, but also on the size of the stellar surface ($\propto R_*^2$). The transformed radius $R_{\rm t}$ \citep{Schmutz1989}, defined as 

\begin{equation}
 R_\text{t} = R_* \left[ \frac{v_\infty}{2500\,{\rm km}\,{\rm s}^{-1}\,}  \middle/  
 \frac{\dot{M} \sqrt{D}}{10^{-4}\,M_\odot\,{\rm yr}^{-1}}  \right]^{2/3},
\label{eq:Rt}
\end{equation}
provides a useful parameter that preserves the strength of emission recombination lines in the model. Models with the same values of $T_*$, $R_{\rm t}$, and chemical abundances would exhibit emission lines of comparable equivalent width, irrespective of $D$, $\dot{M}$, $\varv_\infty$, and $R_*$.

To estimate the optical light ratio in the visual between the two components, we follow \citet{Tehrani2019} and rely on the mass ratio derived in Sect.\,\ref{subsec:RVmeas}. \citet{Graefener2011} provided mass-luminosity relations for homogeneous stars. Given the large masses of the components of \starname~and their large convective cores, the assumption of homogeneity should be reasonable. The luminosities further depend on the hydrogen mass fractions $X_{\rm H}$ of the stars. Fixing $X_{\rm H}$ to the values derived in our analysis (see below), we find that the luminosity of the hotter primary should be larger by about 0.05\,dex. Alternatively, if we assume that both stars comprise entirely of helium (i.e., the mass of the outer hydrogen layer is negligible), the luminosity difference should be 0.03\,dex, which is comparable. Such luminosity differences are obtained when assuming an optical light contribution of $44\%$ for the hotter primary and $56\%$ for the cooler secondary, which we adopt here.

The effective temperatures are determined from the balance of lines belonging to N\,{\sc iii, iv, v}. The wind parameters $R_{\rm t}$ and $v_\infty$ are determined from the strengths and shapes of recombination lines and P-Cygni lines. The mass fractions $X_{\rm N}, X_{\rm H}, $ and $X_{\rm C}$ are determined from the overall strengths of H, He, N, and C lines. The luminosities and reddening are derived by comparing the sum of the SEDs of both components to available photometry and flux-calibrated FUSE and IUE spectra. The reddening comprises two laws: a Galactic foreground contribution following \citet{Cardelli1989} with a reddening of $E_{B-V, {\rm Gal}} = 0.03\,$mag and  $R_V = 3.1\,$, combined with the reddening law derived by \citet{Howarth1983} for the LMC extinction. Our final provided reddening value represents the sum of both contributions: \mbox{$E_{B-V} =  E_{B-V, {\rm Gal}} + E_{B-V, {\rm LMC}} = 0.03 + 0.17 = 0.20\,$mag}, 

\begin{figure}
\centering
\includegraphics[width=.5\textwidth]{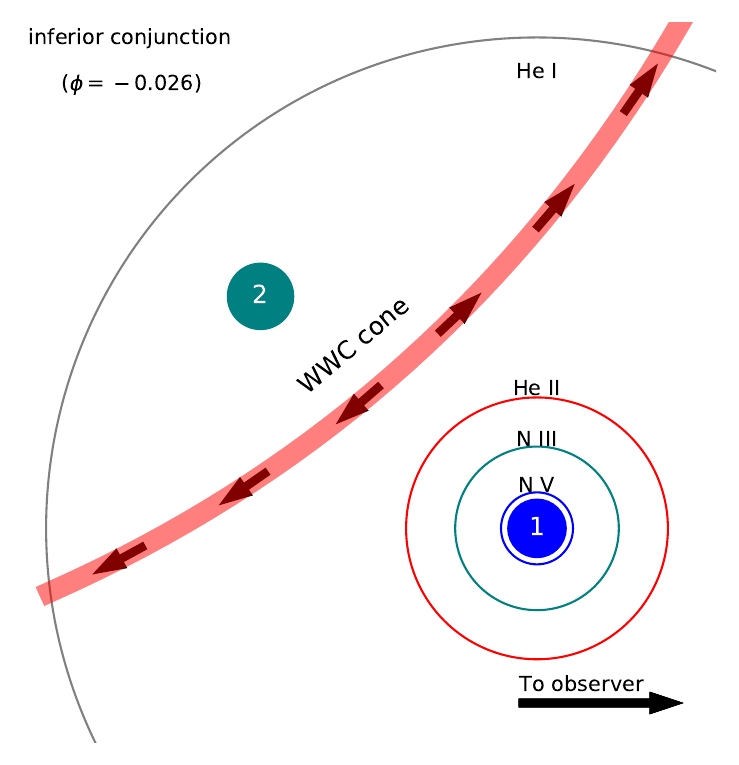}
\caption{Schematic of \starname~as seen during primary inferior conjunction ($\phi \approx -0.026$). The primary (filled blue circle) and secondary (filled teal circle) are plotted to scale with their relative separation. The thick red line indicates the wind-wind collision (WWC) cone. Marked are the line-forming regions of several diagnostic lines for the primary star, as calculated from our tailored model atmosphere (Sect.\,\ref{subsec:nonLTE}). The line-forming regions of the secondary star are comparable.} \label{fig:sketch}
\end{figure}

\begin{figure}
\centering
\includegraphics[width=0.48\textwidth]{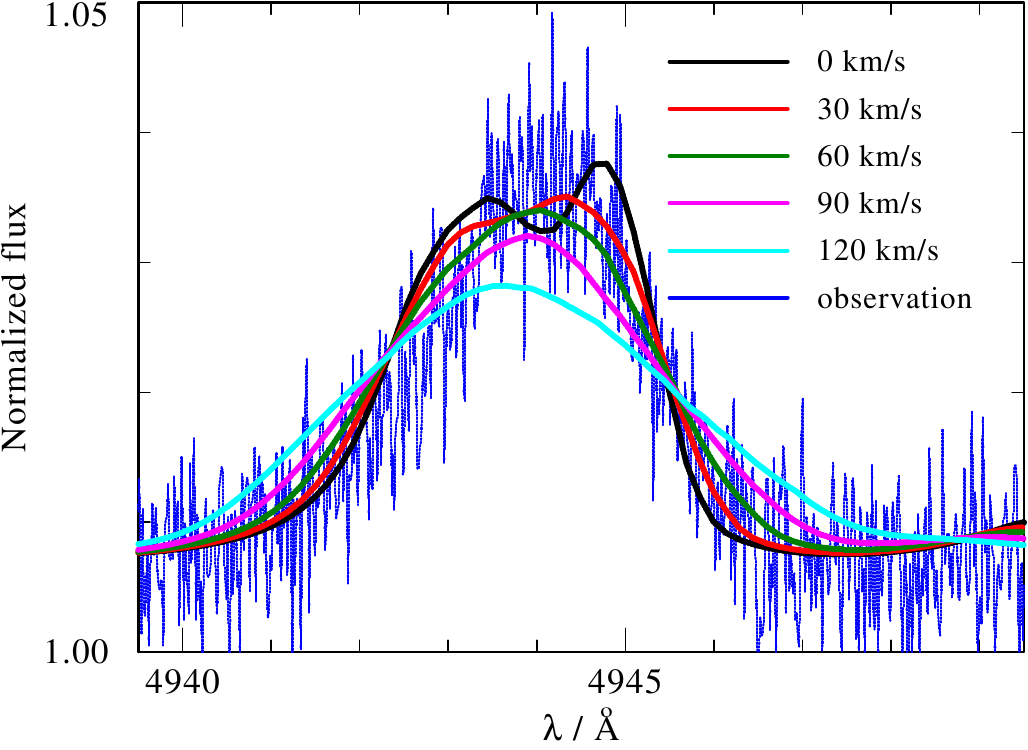}\\
\caption{Comparison between an observed spectrum of the \NVred~multiplet and model spectra for the primary calculated with. the parameters in Table\,\ref{tab:parameters}, but with various $v \sin i$ values (see legend). The calculation is performed using a 3D integration of the formal integral \citep{Shenar2014}.
}
\label{fig:vsini}
\end{figure}

A comparison between the final models and observations is shown in Fig.\,\ref{fig:specan}, and our derived parameters are given in Table\,\ref{tab:parameters}.
The overall global fit generally reproduces the observations well, with a few exceptions. First, the strength of the He\,{\sc i} lines is not reproduced. Reproducing them consistently appears to come at a cost of reproducing diagnostic lines such as \NVblue. Since He\,{\sc i} lines form far out beyond the WWC zone, as illustrated in Fig.\,\ref{fig:sketch}, the strengths of He\,{\sc i} lines may strongly deviate from our  model spectra. We therefore do not consider He\,{\sc i} lines in our fit. Second, assuming the baseline LMC abundance adopted here is adequate, we are not capable of reproducing the desaturated line profiles of the \PVres~resonance doublet, which also form in regions that could be highly impacted by WWCs. This is a known problem that is also observed in the analysis of OB-type stars \citep{Fullerton2006}, and is typically thought to be related to clumpiness and porosity in the wind. Increasing the value of the clumping factor $D$ could improve the discrepancy. However, having calculated models with values up to $D=100$, we find that these lines remain saturated in both the primary and the secondary. We were able to reproduce the de-saturation when accounting for optically thick clumps in the formal integration using the macroclumping formalism (\citealt{Oskinova2007, Surlan2013}, Hawcroft et al.\ 2021, submitted). However, this formalism introduces three additional free parameters, and degrades the quality of the fit in other spectral lines. We therefore avoid including macroclumping in our final fit.

Our derived values for the physical parameters of both components are in broad agreement with similar objects \citep[e.g.,][]{Shenar2017b, Tehrani2019, Bestenlehner2020}. The lower derived value of $X_{\rm H}$ in the primary is a consequence of the ratio of pure helium lines to H+He lines, which is slightly larger for the primary in the disentangled spectra (Fig.\,\ref{fig:disspec}), but should be taken with caution in light of potential contamination of line-profile variability on the disentangled spectra. 

Despite the lack of absorption lines, we can provide realistic estimates for the projected rotational velocity $\varv \sin i$ using emission lines that form  close to the stellar surface, such as the \NVred~and \NIV~lines (see Fig.\,\ref{fig:sketch}). To account for rotation in an expanding atmosphere, we utilise the 3D integration module in PoWR when calculating the formal integral \citep{Shenar2014}. A comparison between various $\varv \sin i$ calculations for the primary component and observations is shown in Fig.\,\ref{fig:vsini}, focusing on the \NVred~line. Evidently, $\varv \sin i \approx 60\,$\kms~best reproduces the observed profile. However, there is a significant degeneracy between the adopted microturbulence $\xi$ and $\varv \sin i$, which prevents us from providing actual estimates for $\varv \sin i$. Nevertheless, upper limits can be obtained. For the secondary component, the \NIV~line was used. These upper limits are provided in Table\,\ref{tab:parameters}.

The uncertainties on $T_*, R_{\rm t}, \log L, E_{B-V}, X_{\rm H}, X_{\rm C}, X_{\rm N}, X_{\rm O}$ are order-of-magnitude estimates based on the sensitivity of the fit to these parameters, as explored by the few hundreds of models calculated here. The errors on the dependent variables (e.g., the stellar radii) follow from error propagation.

\subsection{Wind-wind collision}\label{subsec:WWC}

When two dense winds in a massive binary collide, they form a wind-wind collision (WWC) cone, with its tip at the region where the dynamical pressures of both outflows equalize \citep{Stevens1992}. As the material flows along the WWC cone, it cools down and emits light that can be seen as photometric excess, from the X-ray regime \citep[e.g.,][]{Cherepashchuk1976, Pollock1987} down to the infrared and radio \citep{Williams1997}. Furthermore, as the plasma recombines, WWC excess emission can be seen specifically in recombination lines \citep[e.g.][]{Rauw1999, Marchenko2003, Shenar2017b}.

\begin{figure}
\centering
\begin{tabular}{cc}
\includegraphics[width=0.23\textwidth]{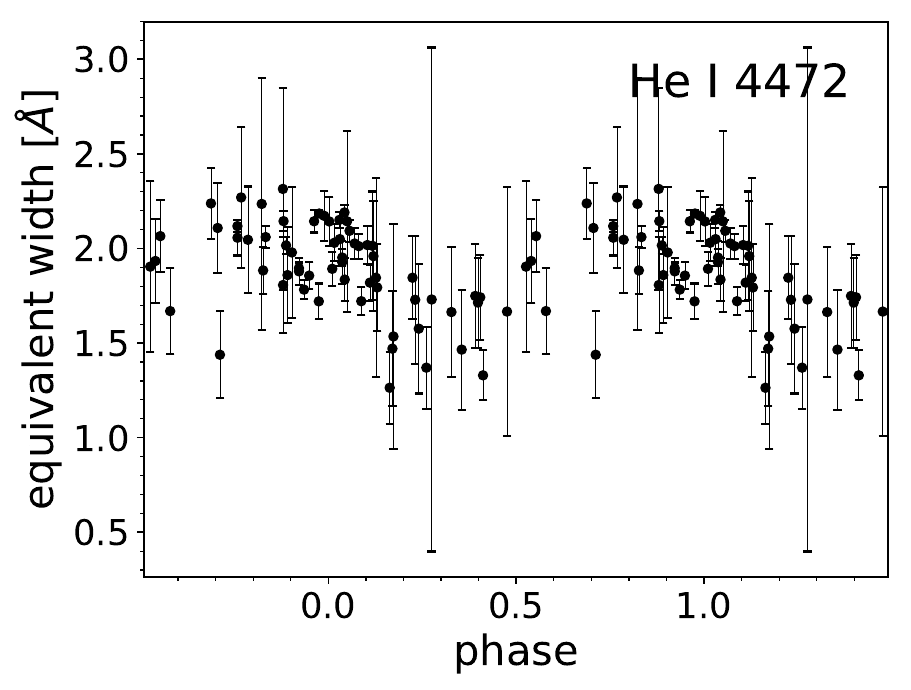} & 
\includegraphics[width=0.23\textwidth]{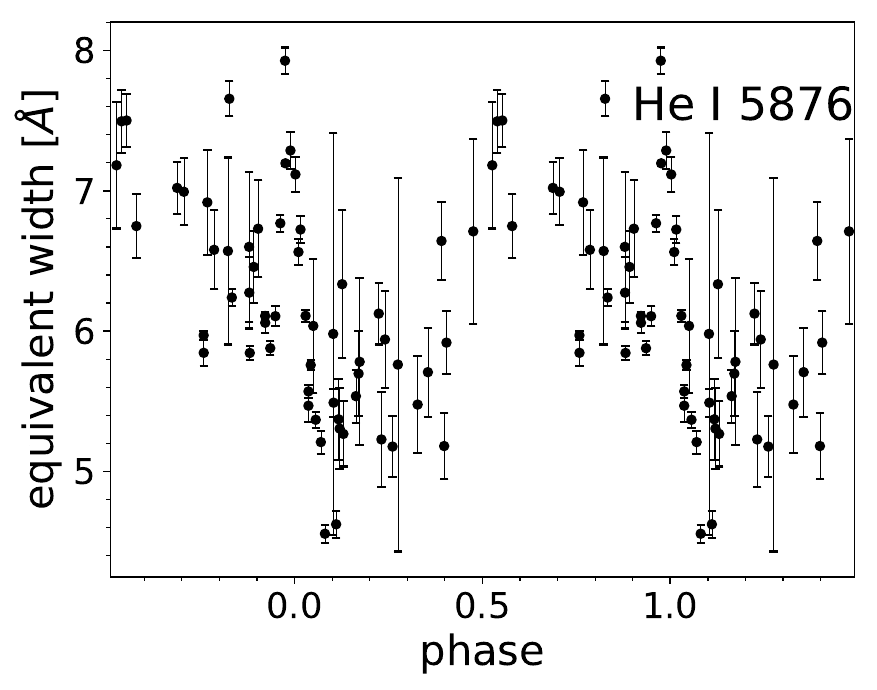} \\ 
\includegraphics[width=0.23\textwidth]{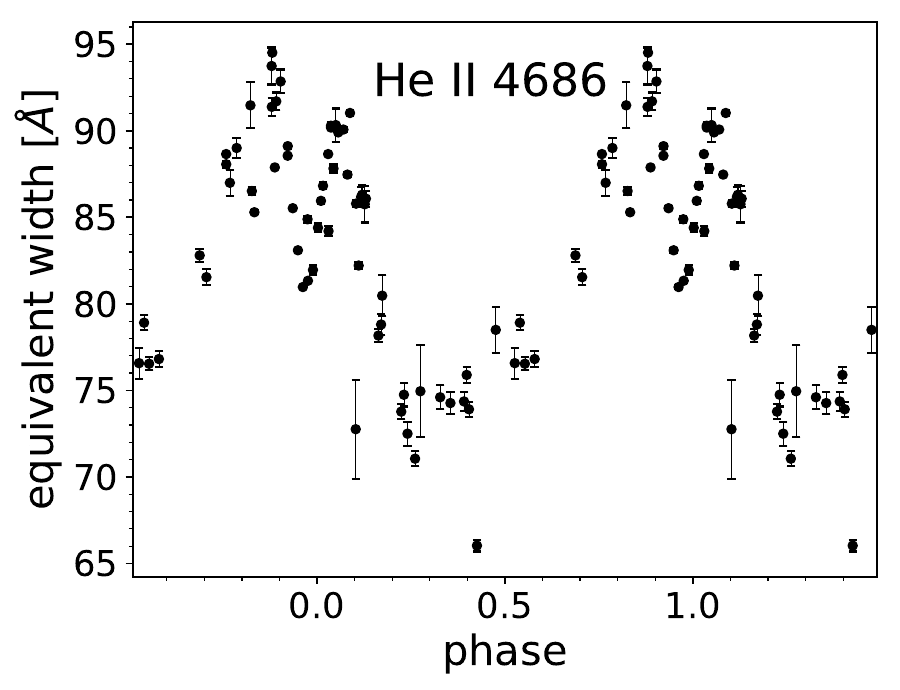} & 
\includegraphics[width=0.23\textwidth]{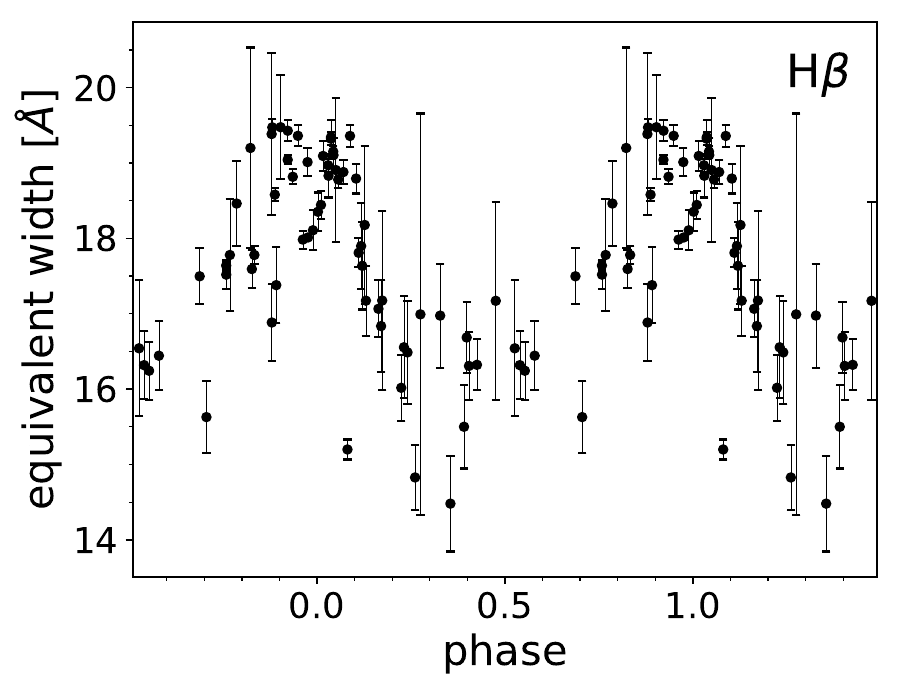} \\ 
\includegraphics[width=0.23\textwidth]{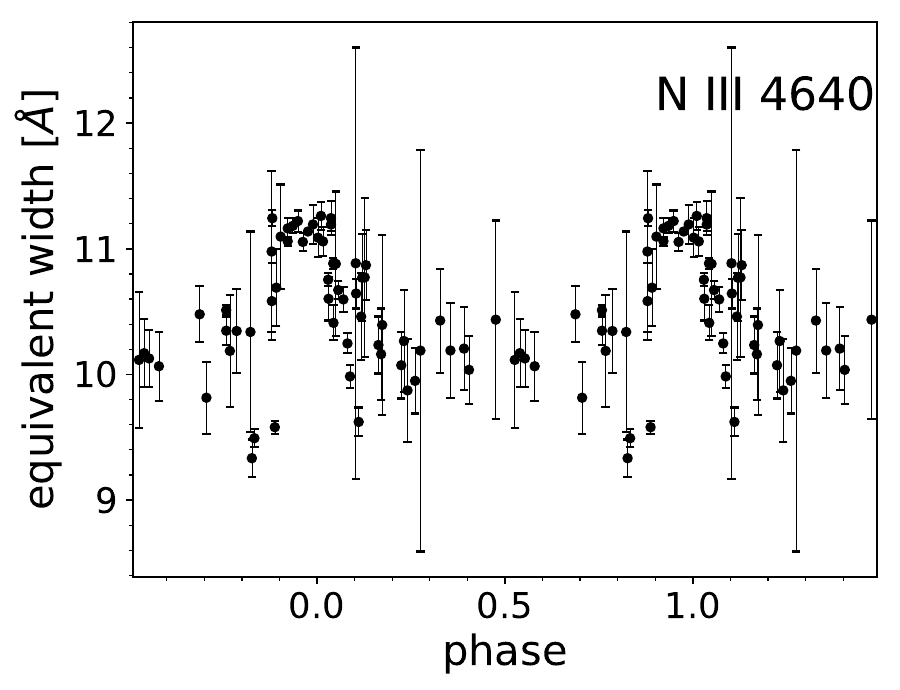} & 
\includegraphics[width=0.23\textwidth]{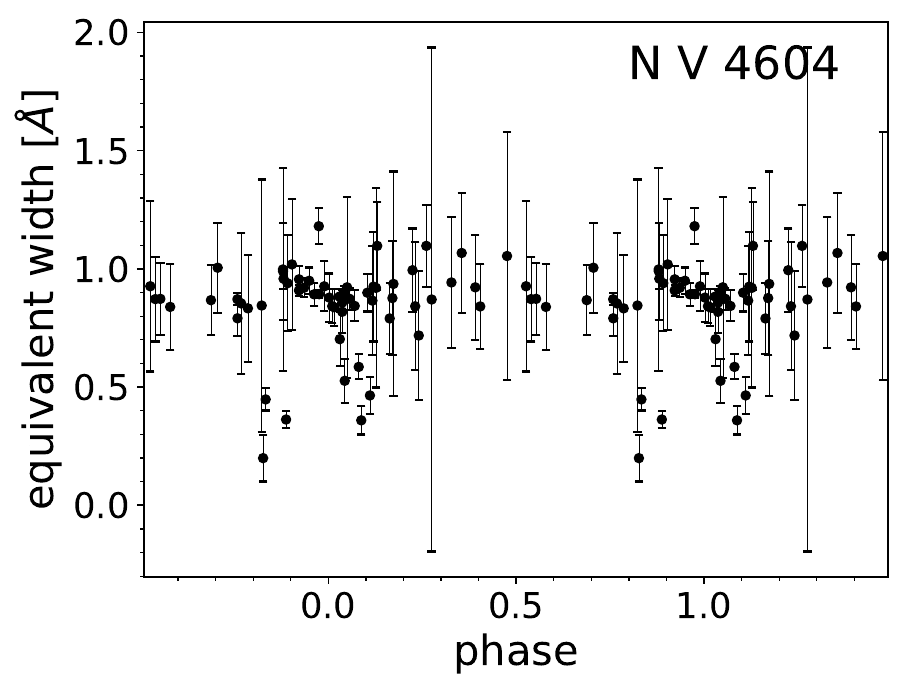} \\ 
\end{tabular}
\caption{EW variations with phase for He\,{\sc i}\,$\lambda 4472$ (4465--4490\,\AA), He\,{\sc i}\,$\lambda 5876$ (5875--5900\,\AA), He\,{\sc ii}\,$\lambda 4686$ (4670--4720\,\AA), H$\beta$ (4840--4890\,\AA), \NIII~ (4620--4650\,\AA), and \NVblue~(4595--4615\,\AA). Only the red part of the He\,{\sc i} lines is integrated to avoid the excessive blue-shifted absorption, which we interpret as line-of-sight absorption by the shock cone
}
\label{fig:WWC_EWs}
\end{figure}

To study whether excess WWC emission may influence the data, we show in Fig.\,\ref{fig:WWC_EWs} the equivalent widths (EW) of several diagnostic spectral lines as a function of orbital phase. It is apparent that almost all lines show excess of flux close to periastron, with the exception of the \NVblue~doublet. In fact, the WWC emission may increase the line flux by up to $\approx 100\%$, but it is absorbed by the stellar winds close to periastron. This can be seen, for example, in the EW change of the \HeII~line (Fig.\,\ref{fig:WWC_EWs}), which increases rapidly towards periastron, but then drops at periastron. Since the emission lines contribute 5-10\% to the visual flux, such a variation in their strengths can lead to changes in the visual light-curve of the order of $\approx 10\,$\%, as we indeed observe (Sect.\,\ref{subsec:LC}).

% Analysis of the WWC features enables in principle the estimation of dynamical and geometric parameters of the system, including the orbital inclination $i$. Unfortunately, there is a large degree of subjectivity in the analysis procedure. We therefore avoid relying on this analysis to recover $i$, but present our  \cite{Luehrs1997} developed a formalism to model the WWC excess features, later simplified  by \citet{Hill2000}. We perform this analysis on the He\,{\sc ii}\,$4472$ line, which appears to form primarily in the WWC region. However, the  One may describe the WWC excess features with two variables that are relatively straight forward to measure: the radial velocity (${\rm RV}_{\rm ex}$) and the full-width half-maximum (${\rm FWHM}_{\rm ex}$) of the feature. As the WWC cone shifts position and orientation, ${\rm RV}_{\rm ex}$ and ${\rm FWHM}_{\rm ex}$ change. Their change as a function of the true anomaly of the orbit $\nu$ can be written as:

\subsection{Visual light-curve}
\label{subsec:LC}

\begin{figure}
\centering
\includegraphics[width=0.5\textwidth]{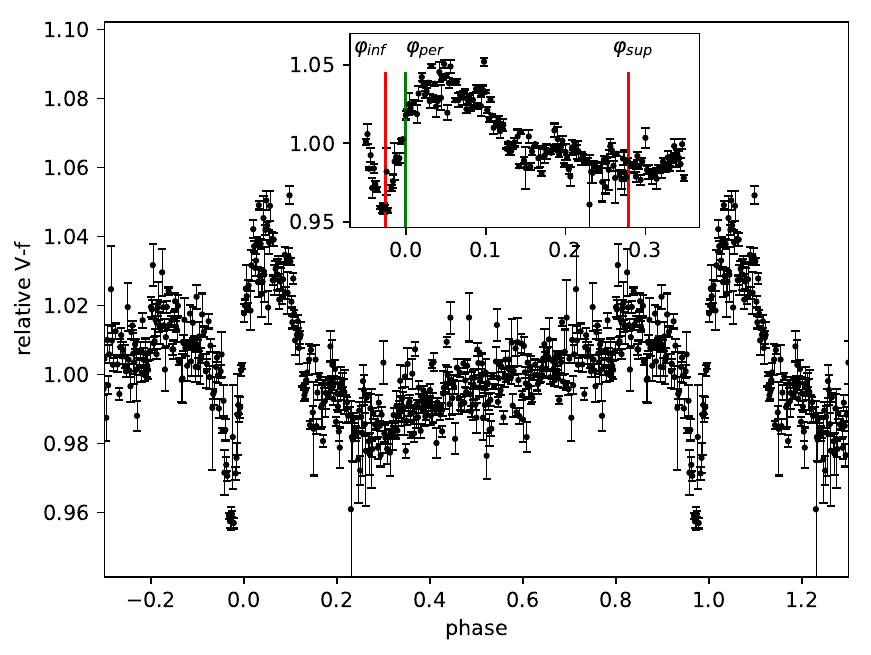}
\caption{ASAS-SN light-curve of \starname~phased with the ephemeris given in Table\,\ref{tab:parameters}. Marked in the inset are the phases of inferior and superior conjunction (hotter primary in front and behind the cooler secondary, respectively) and periastron passage.
}
\label{fig:ASASLC}
\end{figure}

The ASAS-SN optical light-curve of \starname~shows a clear  periodic signature (see Fig.\,\ref{fig:ASASLC}), which covers roughly 22 orbital cycles. The same signature is seen in our extracted TESS light-curve, which covers 1.5 orbital cycles (Appendix\,\ref{appendix:TESS}).
To derive the period, we combined the ASAS-SN and TESS data, which together cover a time base of 6.5\,yr. We then conducted a Period search with Phase Dispersion Minimization \citep[pwkit pdm,][]{Stellingwerf1978, Schwarzenberg-Czerny1997} to derive the period of \starname~of $P = 74.2074\pm0.0043$\,d (Table\,\ref{tab:parameters}). The period is consistent with that found from fitting the RV curves, but given the sharp minimum observed in the light-curve, we fix the period to that found from the light-curve.

% However, the time span of the TESS light-curve only covers 1.5 orbital periods

% The amplitude of this eclipse is quite sensitive to the choice of aperture mask and background subtraction in the TESS reduction procedure. Hence 'third light' is quite substantial in the TESS light-curve. Hence the TESS data serve as a good sanity check and allow us to refine the orbital period, but we do not include the TESS data in modelling the WWC eclipse

Figure\,\ref{fig:ASASLC} shows the ASAS-SN light-curve of \starname~phased with the ephemeris given in Table\,\ref{tab:parameters} and binned at $\Delta \phi = 0.002$. The light-curve mimics that of heartbeat stars \citep{Thompson2012} --  a class of eccentric binaries that show rapid periodic variations in their light-curve, typically associated with tidal distortion of the star(s) during periastron passage. 
We attempted to model the ASAS-SN light-curve using the Physics of Eclipsing Binaries (PHOEBE) light-curve modelling tool \citep{Prsa2016}, using our derived stellar and orbital parameters as input.  Our PHOEBE model is shown in Appendix\,\ref{appendix:PHOEBE}.
However, the observed amplitude of the "heartbeat" signal is roughly 100 times larger than obtained in the model, and the shape was not well reproduced, suggesting that a different mechanism is responsible for the observed behaviour. As we show below, the light-curve can be explained by a combination of  phase-dependent excess emission and eclipses.

To model the light-curve, we construct a hybrid model containing two physical ingredients. 
First, as discussed in Sect.\,\ref{subsec:WWC}, recombination photons stemming from the WWC region can cause flux changes of the order of several percent. This is confirmed from the variation of integrated line-flux in our spectra, and is observed in other WR
(e.g., \object{$\gamma^2$\,Vel} -- \citealt{Richardson2017}). For large separations where adiabatic conditions prevail, this emission is expected to grow roughly as $E_{\rm WWC} \propto 1/\mathcal{D}(\phi)$ reaching a maximum at periastron ($\phi=0$), where $\mathcal{D}(\phi)$ is the instantaneous separation between the stars \citep{Usov1992}. We model the excess emission stemming in WWC as

\begin{equation}
    E_{\rm WWC} (\phi) = A_{\rm WWC}\cdot \left(\frac{\mathcal{D}_{\rm min}}{\mathcal{D}(\phi)}\right)^{\gamma_{\rm WWC}} =  A_{\rm WWC}\cdot \left(\frac{1 + e\,\cos \nu}{1 + e}\right)^{\gamma_{\rm WWC}},
\end{equation}
where $\nu(\phi)$ is the true anomaly, $A_{\rm WWC}$ is the maximum excess at periastron, and   $\gamma_{\rm WWC}$ is a number of the order of unity. $A_{\rm WWC}$ and $\gamma_{\rm WWC}$ are treated as free fitting parameters.  

Second, we account for eclipses in the system. Given its shape and strength,
the flux minimum observed precisely at inferior conjunction ($\phi = -0.026$, primary in front of secondary) is not an ordinary eclipse, but rather a "wind eclipse". As one star is behind the other, the light from the eclipsed star is scattered off the free electrons in the wind of the eclipsing star \citep{Lamontagne1996, St-Louis2005}. \citet{Lamontagne1996} developed a model for the case of O+WR binaries in circular orbits involving a single eclipse of the O star by the WR wind. Here, we extend this model to elliptical orbits by substituting the angle $2 \pi \phi$ with the angle\footnote{Sometimes also denoted as $\lambda$} $\varpi = \nu + \omega + \pi/2$ (defined to be 0 when the primary eclipses the secondary, i.e., at $\nu = -\omega - \pi/2$), and substituting the constant separation $a$ in the case of a circular orbit with $\mathcal{D}(\varpi)$. We further consider the fact that both stars have significant winds and are eclipsing each other by adding the corresponding terms for the companion. We adopt the terminal velocities and stellar radii $R_*$ given in Table\,\ref{tab:parameters}. 
Aside from the orbital parameters $P, e, \omega, T_0$, 
the hybrid model contains six fitting parameters: $A_{\rm WWC}$, $\gamma_{\rm WWC}$, $\dot{M}_1$, $\dot{M}_2$, $i$, and the visual light ratio $f_1/f_2 (V)$. However, we cannot fit the light ratio simultaneously to both $\dot{M}_1, \dot{M}_2$ as the model becomes degenerate. Therefore, we fix the light ratio to the value estimated in Sect.\,\ref{subsec:nonLTE} ($f_1/f_2 = 0.79$), but discuss the impact of changing this value in Appendix\,\ref{appendix:systematics}.

The model relies on the following assumptions: (i) The stars can be approximated as point sources. With the derived geometry of \starname, this should be a fairly good approximation (however, see discussion in Appendix\,\ref{appendix:systematics}). (ii) The optical depth along the line-of-sight is dominated by Thomson scattering off free electrons. This assumption is verified by our model atmospheres, where electron scattering contributes $\approx 80\%$ to the total opacity in the outer layers.  (iii) The scattering is optically thin. This is verified by our model, where the optical depths typically remain well below unity. (iv) The wind velocity field takes the form of a $\beta$-law. Analytical solutions are provided by \citet{Lamontagne1996} for the $\beta=0$ (constant velocity) and $\beta=1$, but arbitrary values of $\beta$ may be assumed if the integration is performed numerically. For now, we adopt $\beta=1$, but explore $\beta=2$ in Appendices\,\ref{appendix:systematics} and \ref{appendix:beta}.  (v) The wind is fully ionized, which yields the number of electrons per baryon $\alpha \simeq (1+ X_{\rm H})/2$, resulting in $\alpha_1 = 0.68$ and $\alpha_2 = 0.7$. Our model atmospheres suggest that the assumption of fully-ionized winds holds reasonably well for the primary, but breaks  down at a few stellar radii above the stellar surface for the secondary. For now, we fix $\alpha$ as constant, but discuss this assumption in Appendix\,\ref{appendix:systematics}. (vi) Both stellar winds are spherically symmetric and unperturbed by each other, and no additional opacity sources are present. We discuss the validity of this assumption in Appendix\,\ref{appendix:systematics}.

% \begin{figure}
% \centering
% \includegraphics[width=0.5\textwidth]{Wind_eclipse_fullfit_R144.pdf}
% \caption{Comparison between the observed ASAS-SN light-curve and our hybrid model consisting of double wind-eclipse + WWC excess emission, following \citet{Lamontagne1996}. See text for details
% }
% \label{fig:hybridModel}
% \end{figure}

\begin{figure}
\centering
\includegraphics[width=0.5\textwidth]{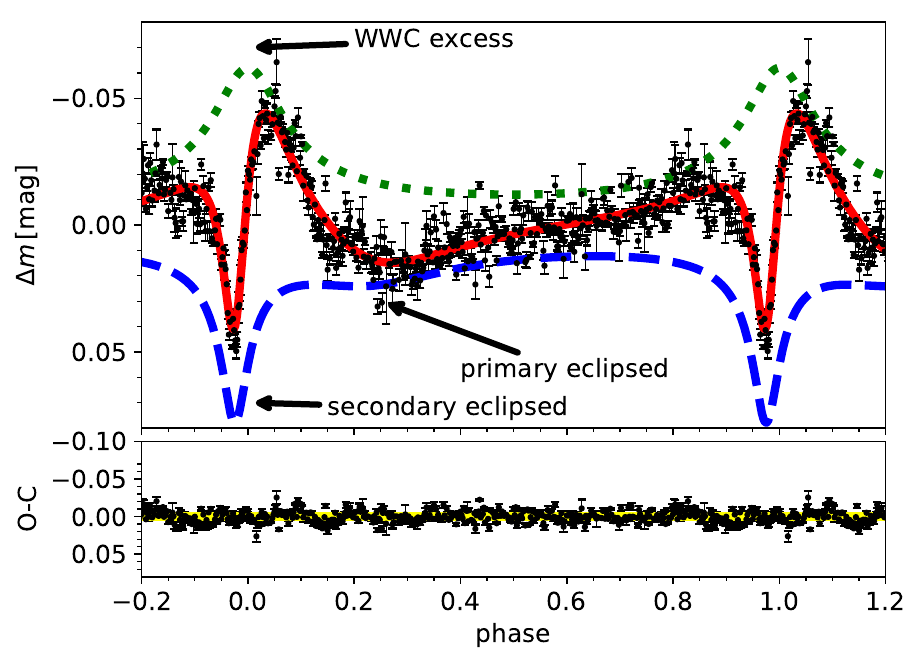}
\caption{{\it Upper panel:} Comparison between the observed ASAS-SN light-curve and our best-fitting hybrid light-curve model (red solid line) consisting of double wind-eclipse (blue dashed line) + WWC excess emission (green dotted line), which are multiplied by 0.5 for clarity.  {\it Lower panel:} Residuals between observation and model (O-C).
}
\label{fig:hybridModel}
\end{figure}

The ASAS-SN light-curve is fitted simultaneously with the RV curves derived in Sect.\,\ref{subsec:RVmeas}.
We refrain from combining the TESS data with the ASAS-SN data in the fit. This is both because the time span of the TESS light-curve only covers 1.5 orbital periods (i.e., only one wind-eclipse is seen), and also because the amplitude of the eclipse is sensitive to the choice of aperture mask and background subtraction in the TESS reduction procedure due to the relatively large pixel size of TESS. However, the TESS data allow for a refinement of the orbital period, and provide an encouraging sanity check for our model. A comparison between our model and the TESS data is presented in the Appendix\,\ref{appendix:TESS}.  A comparison between our model and the ASAS-SN data is shown in Fig.\,\ref{fig:hybridModel}. The derived parameters are \mbox{$A_{\rm WWC} = 0.12\pm0.01$}, $\gamma_{\rm WWC} = 1.5\pm0.1$, $\log \dot{M}_1 = -4.29\pm0.05\,$\smy, $\log \dot{M}_2 = -4.42\pm0.05\,$\smy, and $i=60.4^\circ$ (given also in Table\,\ref{tab:LCmodel}).

\begin{figure}
\centering
\includegraphics[width=0.5\textwidth]{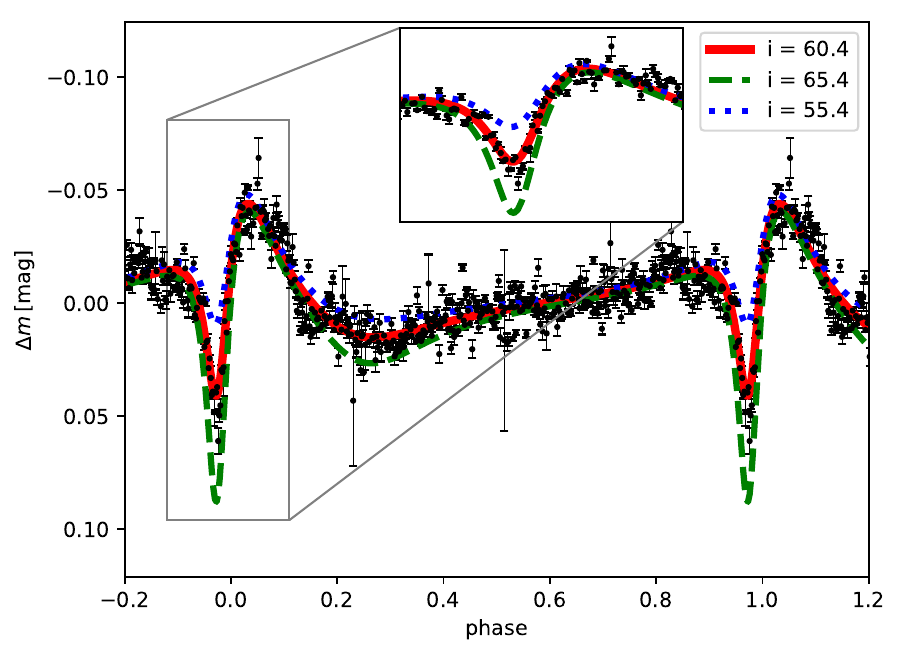}
\caption{Same as Fig.\,\ref{fig:hybridModel}, but showing the impact of changing the inclination by $\pm 5^\circ$ while fixing all other parameters (green and blue lines, see legend).
}
\label{fig:hybridModelInc}
\end{figure}

The agreement between the ASAS-SN data and our analytical model is excellent, with a formal reduced $\chi^2$ for the RV curves + light-curve combined of $\chi^2 = 0.7$. As this indicates that the formal errors on the photometry are overestimated, we rescale them with a constant factor (0.82) to obtain $\chi^2$ values of the order of unity, which ensures that the relative weighting between the photometric and RV measurements is adequate\footnote{The conclusion that the noise is overestimated assumes that the model is adequate and the noise is Gaussian. Nevertheless, we note that we get very similar results when no correction is applied to the errors.}.  Figure \ref{fig:hybridModel} also shows the residuals between observation and computation (O-C). No clear discrepancies are seen. However, the O-C plot reveals a low-frequency modulation, also observed in other WR stars \citep[e.g.][]{Ramiaramanantsoa2019}. However, a periodogram of the residuals did not reveal any signficant frequency peaks.

While it is not immediately evident, the light-curve is impacted by both eclipses. The eclipse at primary superior conjunction ($\phi = 0.28$, secondary is in front) is responsible for the rapid decrease of flux after periastron and to the slight discontinuity observed at $\phi = 0.28$. It is wider than the eclipse at $\phi = -0.026$ due to the longer time spent at apastron, and it is weaker primarily due to the larger separation between the two components, resulting in a larger impact parameter and hence lower column density.

The derived parameters are in very good agreement with expectation. The mass-loss rates, $\log \dot{M}_1 = -4.29\pm0.05\,$[\smy] and $\log \dot{M}_2 = -4.42\pm0.05\,$[\smy], agree within $1\sigma$ with those derived from spectroscopy. The WWC excess reaches a maximum of $\approx \pm10$\%, which is roughly consistent with the flux changes in recombination lines (Fig.\,\ref{fig:WWC_EWs},  Sect.\,\ref{subsec:WWC}).  The exponent of $\gamma_{\rm WWC} = 1.5\pm0.1$ is in the accepted range of 1-2 \citep[e.g.][]{Usov1992, Schnurr2009, Shenar2017b}. 

Importantly, the  modelling of the light-curve enables us to constrain the orbital inclination. In fact, the model is extremely sensitive to the inclination. This is illustrated in Fig.\,\ref{fig:hybridModelInc}, where we show the impact of changing  $i$ by $\pm5^\circ$ when fixing all other parameters. Our fit implies an inclination of $i = 60.4\pm1.5^\circ$. This, in turns, implies dynamical masses of $74\pm4\,M_\odot$ and $69\pm4\,M_\odot$ for the primary and secondary, respectively. The validity of our model and an investigation for possible systematic errors are thoroughly discussed in the Appendix\,\ref{appendix:systematics}.

\subsection{Polarimetric analysis}
\label{subsec:Polan}

\begin{figure}
\centering
\begin{tabular}{c}
\includegraphics[width=0.475\textwidth]{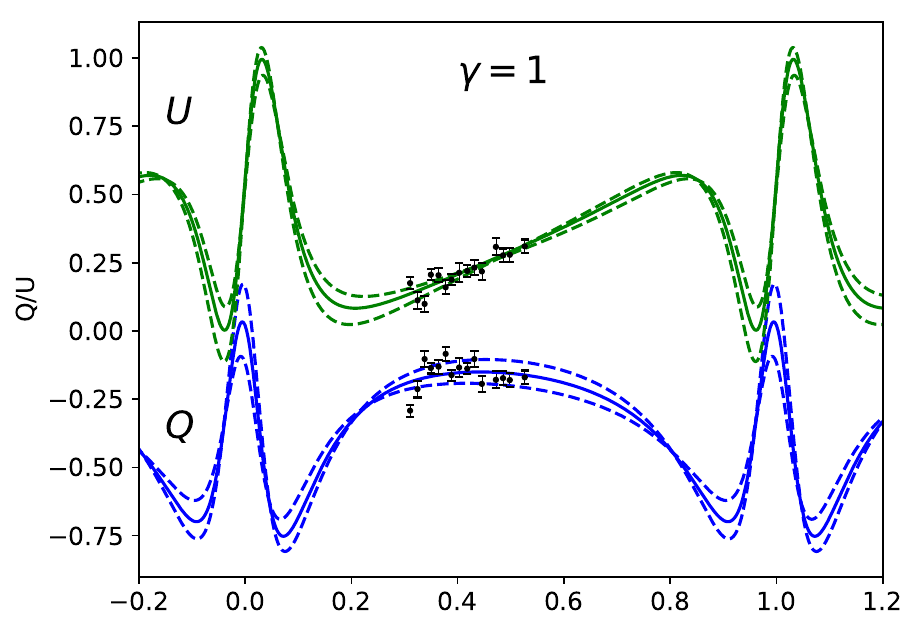} \\
\includegraphics[width=0.475\textwidth]{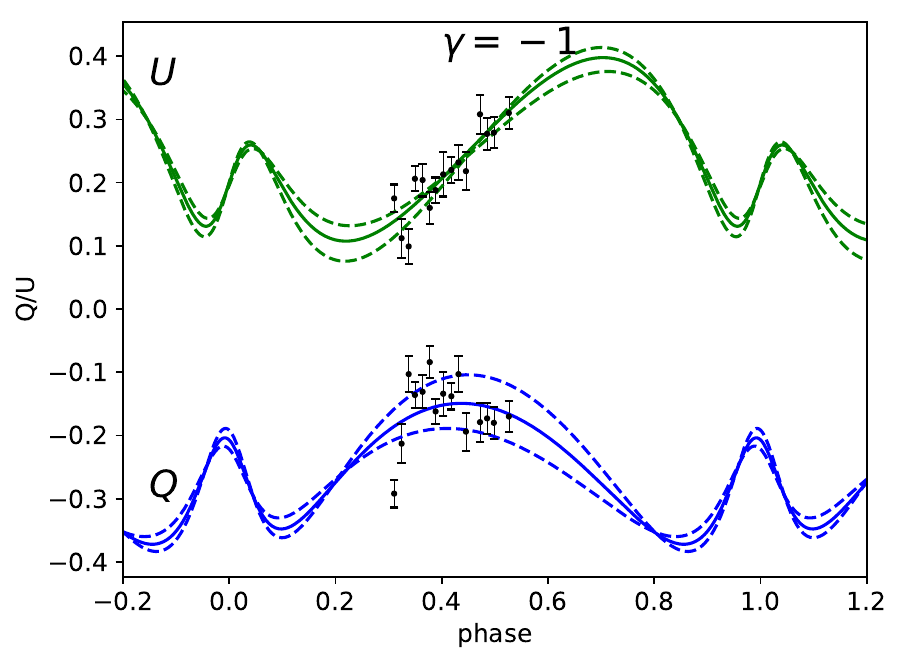} 
\end{tabular}
\caption{Comparison between the observed and modelled Stokes parameters $Q$ (blue line) and $U$ (green line) for two distinct exponents $\gamma$. For $\gamma = 1$ (upper panel), we obtain $\Omega = 120\pm11^\circ$, $\tau_* = 0.22$, $Q_0 = -0.24\pm0.02$, $U_0 = 0.16\pm0.02$. For $\gamma=-1$ (lower panel), we obtain  $\Omega = 125\pm11^\circ$, $\tau_* = 0.10\pm0.02$, $Q_0 = -0.23\pm0.02$, $U_0 = 0.15\pm0.02$. Dashed lines bounding the curves show the impact of changing the inclination by $\pm 10^\circ$. See text for details.
}
\label{fig:Polfig}
\end{figure}

An additional observational constraint that is sensitive to the inclination is provided by polarimetry. \starname~was studied using spectropolarimetry by \citet{Vink2017}, who reported mild depolarization in recombination lines ("the line effect"), presumably due to the binary nature of \starname. Time-dependent polarimetry can yield important constraints on the geometry of the system, including the orbital inclination.  When the light of a star is scattered off the free electrons in the wind of its companion, a net polarization can be observed in the Stokes parameters $Q$ and $U$ that varies with time. \citet{Brown1978, Brown1982} developed analytical equations that describe the change in $Q$ and $U$ in elliptical systems where one star possesses a wind\footnote{The equations were later corrected by \citet{Simmons1984}.}. This model has since been implemented in several cases of WR binaries \citep{St-Louis1988, Moffat1998}, including the WR+WR binary \object{R\,145} \citep{Shenar2017b}. The phase change of the Stokes parameters depends not only on the orbital parameters $P, T_0, e, \omega$, and $i$, but also on the longitude of the ascending node $\Omega$. Moreover, the model depends on the density and structure of the scattering medium, which is typically parametrised with two free fitting parameters: an effective optical depth $\tau_*$, and an exponent related to the structure of the wind, $\gamma$. For brevity, we refer to  \citet{Shenar2017b} for a full account of the relevant equations. 

Unfortunately, only a few polarimetric measurements are available for \starname, and they are obtained relatively far from periastron. However, we can still fix most of the parameters based on our previous analyses (cf.\ Table\,\ref{tab:parameters}), leaving only $\Omega$, $\tau_*$, $\gamma$ and two arbitrary offset constants $Q_0$ and $U_0$ as fitting parameters. The fitting procedure, performed using the Python lmfit package, reveals that $\gamma$ and $\tau_*$ cannot be well constrained with the available data, while $\Omega$ can be. Therefore, we provide our preliminary value of $\Omega$ in Table\,\ref{tab:parameters}. In Fig.\,\ref{fig:Polfig}, we show the comparison between the $Q/U$ data and our-best-fitting models for two typical values of $\gamma = -1$ and $1$. The impact of changing the inclination is also shown.

The few data points available, along with their relatively large errors, can accommodate a wide range of inclination angles, and we therefore cannot provide an independent measurement of $i$, which is critical for the estimation of the dynamical masses. Future polarimetric monitoring of the system should is therefore strongly encouraged.

\subsection{X-ray light-curve}
\label{subsec:LCXR}

\begin{figure}
\centering
\includegraphics[width=.5\textwidth]{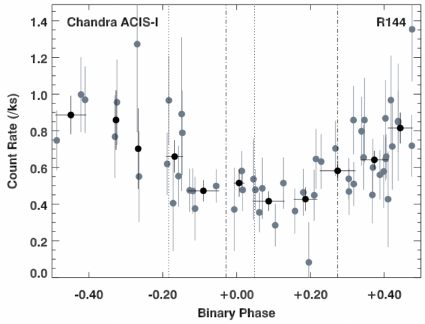}
\caption{X-ray light-curve of \starname, also shown aggregated into adaptive wider phase intervals. See text for details.} \label{fig:XRLC}
\end{figure}

As discussed in Sect.\,\ref{subsec:WWC}, with shock velocities of the order of $\approx 1000\,$\kms, WWCs are expected to lead to substantial X-ray emission \citep[e.g.,][]{Cherepashchuk1976, Stevens1992, Usov1992}. X-ray excess is indeed often, though not always, observed in binaries with two massive components \citep[e.g.,][]{Pollock1987, Corcoran1996, Naze2007, Guerrero2008}. Similarly to the optical emission excess in recombination lines, one would expect the X-ray emission to increase at periastron, where the densities are highest, and decrease inversely proportional to the separation. 

Interestingly, the exact opposite is observed in \starname.
In Fig.\,\ref{fig:XRLC}, we show the T-ReX X-ray light-curve of \starname, folded with the ephemeris given in Table\,\ref{tab:parameters}.  \citet{Tehrani2019PhD} reported \starname~to be a fairly hard X-ray source ($kT = 4.2\pm0.4\,$keV) and reported with an X-ray luminosity of $\log L_{\rm X} = 33.4\,$\ergs.  For such a luminous binary, this is relatively modest. In comparison, the massive WNh+WNh binary \object{Mk\,34}, despite sharing many of the characteristics of \starname~and despite portraying a comparable X-ray hardness ($kT = 3.16\pm0.03\,$keV), exceeds this X-ray luminosity by almost two orders of magnitudes at peak luminosity ($\log L_{\rm X} = 35.3\,$\ergs, \citet{Pollock2018}). Moreover, while the X-ray flux of \object{Mk\,34} increases shortly before periastron (followed by a rapid decrease), it decreases at periastron in the case of \starname. This is in contrast to the pattern observed in the optical light-curve. 

The difference in the X-ray  behaviour between \object{Mk\,34} and \starname~may be related to their different wind properties. The components of \object{Mk\,34} have reported terminal velocities that are roughly twice as large as reported here for \starname, but they also have mass-loss rates that are about three times lower than reported here. Additionally, with an orbital period of $P=155\,$d,  the separation between the components in \object{Mk\,34} is significantly larger than found for \starname. The smaller semi-major axis in \starname~in combination with the stronger winds (compared to Mk\,34) could mean that a substantial amount of X-rays is absorbed by the stellar winds. Moreover, at the winds likely do not reach their terminal value at the collision region. Assuming $\beta=1$ for the velocity law, the wind velocities are roughly 75\% (i.e., $\approx 1000\,$\kms) at collision during periastron, and even lower for larger $\beta$ values. Hydrodynamical simulations will be needed to assess whether these facts are sufficient to explain the two orders-of-magnitude difference observed in their X-ray luminosities. Importantly, \starname~shows that, while detection of strong X-rays is a helpful indicator of multiplicity, the converse does not hold. 

% which are beyond the scope of the current study.

% The fact that the optical emission excess peaks at periastron while the X-ray flux drops could indicate that the velocities of the colliding winds are not fast enough at the time of collision to produce hard X-rays. 

% This is because, at periastron, the collision is only $\approx 5\,R_*$ above the stellar surfaces, where the wind is still being accelerated. With $\beta=1$, roughly $1000\,$\kms are reached, but for higher $\beta$ values substantially lower values are reached. 

\section{Discussion}
\label{sec:discussion}

\subsection{Evolutionary status}
\label{subsec:evolution}

We estimate the evolutionary masses and ages of both components using the Bayesian statistics tool BONNSAI\footnote{ The BONNSAI web-service is available at
www.astro.uni-bonn.de/stars/bonnsai} \citep{Schneider2014}. 
Using a set of input stellar parameters and
their corresponding errors, the tool estimates the current and initial masses along with the age by interpolating between evolutionary tracks calculated at LMC metallicity by \citet{Brott2011} and \citet{Koehler2015} for stars with initial
masses up to 500~\msun\ and over a wide range of initial rotation
velocities. As input parameters, we use the derived values of $T_*$, $\log L$, $X_{\rm H}$, and upper limits on $\varv_{\rm eq}$. However, the BONNSAI tool cannot find a solution that satisfies these three constraints. We therefore advanced by omitting the condition on the bounded rotational velocity.

\begin{figure}
\centering
\includegraphics[width=0.5\textwidth]{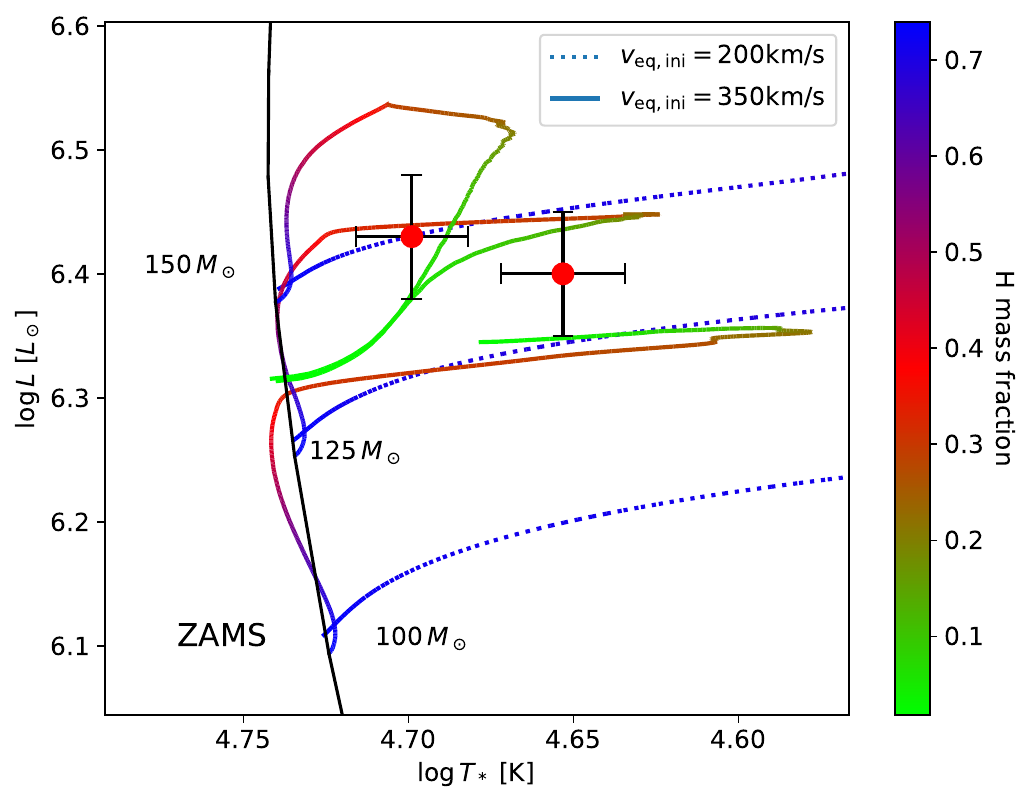}
\caption{HRD positions of the primary and secondary components of \starname, along with evolution tracks calculated by \citet{Koehler2015} for $M_{\rm ini} = 100, 125, $, and 150~\msun\ with an initial rotation of $\varv_{\rm eq, ini} = 200\,$\kms~and 350\,\kms. The color corresponds to the surface hydrogen mass fraction.
}
\label{fig:BonnTracks}
\end{figure}

Our final results are shown in Table\,\ref{tab:parameters}. BONNSAI infers an age of roughly 2\,Myr for both components. The consistent ages suggest that the components of \starname~have not interacted via mass-transfer in the past, as expected from the orbital separation and the significant eccentricity. We derive current masses of $M_{\rm ev, cur} = 110\pm9$ and $100\pm10$~\msun\ and initial masses of $M_{\rm ini} = 130\pm13$ and $119\pm12$~\msun\ for the primary and secondary, respectively (Fig.\,\ref{fig:BonnTracks}). 

% A set of evolution tracks calculated by \citet{Koehler2015} with initial masses $M_{\rm ini} = 100,  125,$ and 150~\msun\ and an initial rotation of $v_{\rm eq, ini} = 200$ and $350\,$\kms~are shown in an HRD in Fig.\,\ref{fig:BonnTracks} along with the derived positions of the binary components of \starname. 

% Comparing the minimum masses $M \sin^3 i$ with the evolutionary masses yields $i = 49.5\pm2.8^\circ$ for the primary and $i=51.4\pm3.1^\circ$ for the secondary, which we simplify to $i=50\pm3^\circ$. This agrees well with our estimate from the WWC analysis (Sect.\,\ref{subsec:WWC}) of $i=49^\circ$. However, the evolutionary masses highly depend on the adopted mixing and mass-loss prescriptions, which are still considered uncertain. Therefore, it would be very helpful to derive the inclination from the light-curve in future work (Fig.\,\ref{fig:ASASLC}), which should yield a much better accuracy than the analysis of WWCs.

\subsubsection{Rotation discrepancy}
\label{subsubsec:rotdis}

While the ages of both components are consistent within their  errors, the Bonn models are only capable of reproducing the observed parameters when assuming high initial rotation rates of $\varv_{\rm eq, ini} \approx 350-400\,$\kms, which tend to chemically homogenize the stars. This in turn results in current rotation velocities of $\varv_{\rm eq, cur} \approx 250\,$\kms, which are substantially larger than the conservative upper limits derived here ($\varv_{\rm eq} \lesssim 100$\,\kms). As is illustrated in Fig.\,\ref{fig:BonnTracks}, models without very high initial rotation are not capable of reproducing the strong He enrichment that we observe.

The mass-loss rates implemented by \citet{Koehler2015} rely on extrapolations from \citet{Vink2001}. However, recent empirical \citep{Bestenlehner2014, Bestenlehner2020} and theoretical studies \citep{Graefener2008, Vink2011, Schneider2018, Bestenlehner2020Mdot, Graefener2021}  suggest that mass-loss rates at the upper-mass end are substantially larger than the original \citet{Vink2001} prescriptions. Indeed, the current mass-loss rates retrieved by the BONNSAI tool are of the order of $\log \dot{M} = -4.8\,$[\smy], more than a factor two lower than obtained empirically in our study. Higher mass-loss rates would result in a more rapid spin-down of the stars, potentially bringing the rotation in agreement with observation. Moreover, higher mass-loss rates would enable the stars to strip themselves off their H-rich layers and hence lead to atmospheric H-depletion without the need of invoking very high rotation to homogenise them. Hence, boosting the mass-loss rate would have a "double" effect in this regard.

\begin{figure}
\centering
\includegraphics[width=0.5\textwidth]{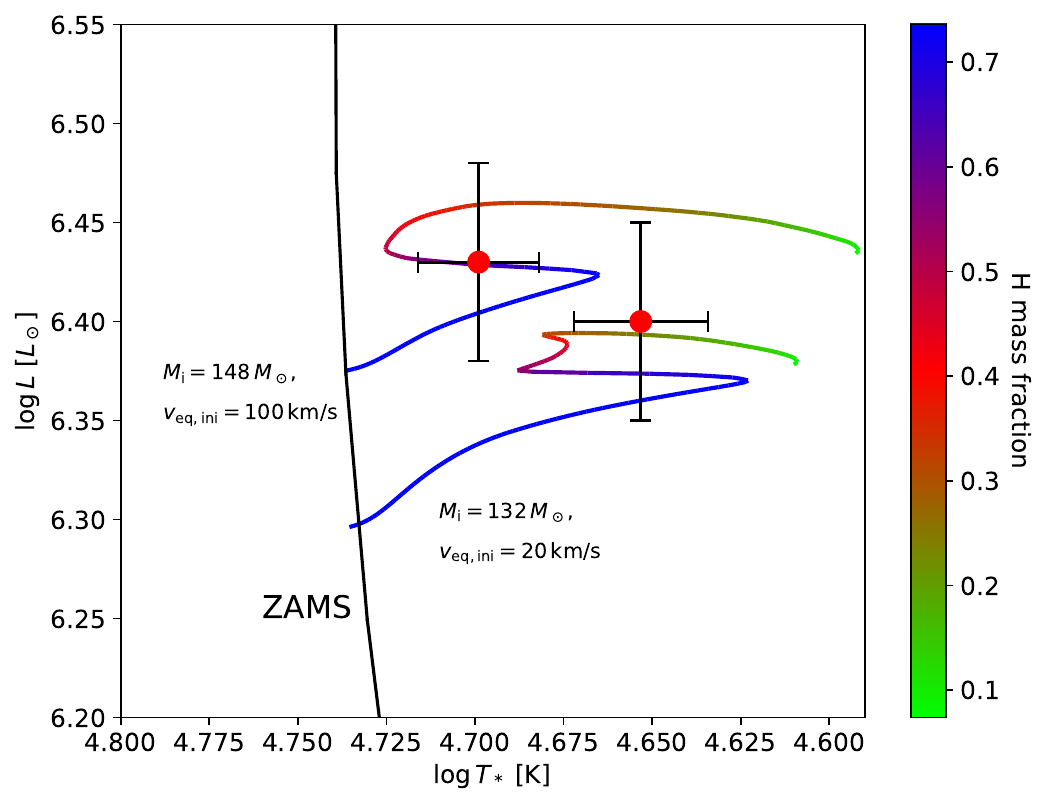}
\caption{As Fig.\,\ref{fig:BonnTracks}, but showing MESA tracks with parameters corresponding to those given in Table\,\ref{tab:parameters}. These tracks are calculated using the  setup described by \citet{Graefener2021}, which includes boosted mass-loss rates.
}
\label{fig:GoetzTracks}
\end{figure}

To explore this further, we compared our results with MESA  evolution code \citep{Paxton2011} calculated by  \citet{Graefener2021}, which include enhanced mass-loss rates. The tracks were integrated into the BONNSAI tool, allowing for a derivation of the most probable initial masses, rotation velocities, and ages. The final results are presented in Table\,\ref{tab:parameters}, and representative tracks are shown in Fig.\,\ref{fig:GoetzTracks}. Evidently, the tracks reproduce the observed HRD positions simultaneously to the observed H-depletion without the need to invoke high rotation. Due to the larger mass-loss rates, the initial masses are larger than those inferred by BONNSAI, but the ages and currernt masses are comparable to those derived with BONNSAI. Since the MESA tracks better represent the properties of the system, and given the evidence for increased mass-loss, we adopt the evolutionary parameters from the MESA tracks and provide them in Table\,\ref{tab:parameters}. 

There are additional mechanisms that could be relevant in the context of rotation and H-depletion. For example, it is also possible that additional mixing mechanisms operates at these very high masses \citep[e.g.,][]{Higgins2019, Jermyn2018, Pedersen2018, Schootemeijer2019, Bowman2019, Gilkis2021}. Moreover, 
tidal interaction at periastron may have also operated in \starname, slowing the components down towards synchronicity. At periastron, the ratio between the separation and diameter of the stars is  $a_1\,(1-e)\,{R_1}^{-1} \approx a_2\,(1-e)\,{R_2}^{-1} \approx 4$. At such ratios, tidal synchronization of the rotational velocities is expected to become important on a nuclear timescale \citep[see table 2 in][]{Zahn1977}. While these mechanisms do not seem to be necessary to explain the observed H-depletion and low rotational velocities, it is possible that they have operated in \starname.

% which may operate both in the convective core \citep{Jermyn2018,  In fact, there is a growing body of evidence that stars exceeding a certain mass tend to evolve quasi-homogeneously regardless of rotation \citep{Vink2017, Shenar2017b, Ramachandran2019}. A possible mechanism could for example be an overshooting scale that grows with stellar mass \citep[e.g.][]{Jermyn2018}. Such a mechanism was recently suggested as a possible origin for the Humphreys-Davidson limit observed in the Magellanic Clouds \citep{Gilkis2021}, though a direct proof of this is still lacking.  Alternatively, envelope-mixing processes that are not induced by rotation (e.g., gravity waves), for which it is not a standard procedure to implement  in evolution codes \citep[e.g.,][]{Pedersen2018, Bowman2019, Tkachenko2020}, could also reduce the observed discrepancy.

% Finally, tidal interaction at periastron may have also operated in \starname, slowing the components down towards synchronicity. At periastron, the ratio between the separation and diameter of the stars is  $a_1\,(1-e)\,{R_1}^{-1} \approx a_2\,(1-e)\,{R_2}^{-1} \approx 4$. At such ratios, tidal synchronization of the rotational velocities is expected to become important on a nuclear timescale \citep[see table 2 in][]{Zahn1977}. As the (psuedo)-synchronous velocity would be $\approx 20$\,\kms, this would slow the components down substantially from potentially much higher initial rotation velocities, and could thus be important in the context of the observed discrepancy.

\subsubsection{Mass discrepancy}
\label{subsubsec:massdis}

The derived evolutionary masses ($M_{\rm ev, cur} \approx 110\,M_\odot, 100\,M_\odot$) disagree with our dynamical masses ($M_{\rm dyn} = 74\pm4\,M_\odot$,  $69\pm4\,M_\odot$), which at least formally are determined to a high degree of accuracy. Interestingly, a similar discrepancy was found by \citet{Shenar2017} for the binary \object{R\,145}, which shares many similarities with \starname.
The combination of high stellar luminosities and relatively low masses imply that the binary components are situated close to the Eddington limit, with Eddington factors of $\Gamma_{\rm e} = 0.78 \pm 0.10$ for both components. Considering the additional pressure from line-driving, it is hard to imagine that the stars should not be strongly inflated at such Eddington factors \citep{Petrovic2006, Graefener2011, Graefener2012, Grassitelli2016, Sanyal2017}. However, the radii and temperatures derived here are consistent with expectation for non-inflated main sequence stars  \citep{Graefener2011}.

This discrepancy has two possible resolutions. On the one hand, it is possible that the errors of one or more of the orbital parameters are underestimated. Of all orbital parameters, the orbital inclination is probably the most critical one here. An inclination of $50^\circ$ -- a mere $10^\circ$ decrease -- would yield masses that are comparable to the evolutionary masses. However, the formal error on the inclination is only $1.5^\circ$, as the eclipse light-curve model is extremely sensitive to $i$ (see Fig.\,\ref{fig:hybridModelInc}). Therefore, if the error is underestimated, it must be systematic in nature. We explored various assumptions in the light-curve model,  and these attempts are described in detail in Appendix\,\ref{appendix:systematics}. We could not isolate a potential source of error to justify a $10^\circ$ difference. Moreover, in the framework of our light-curve model, reducing the inclination down to $50^\circ$ also requires very large mass-loss rates of the order of $\log \dot{M} = -3.9\,$[\smy] to reconcile for the decrease of $i$, which is inconsistent with our spectral analysis.

Taken at face value, the relatively low dynamical masses may suggest an entirely different interpretation for the evolutionary status of the system: the luminosities and masses would be consistent with fundamental structure models when assuming that both stars comprise primarily He in their cores, surrounded by relatively thin layers containing residual hydrogen \citep[e.g.][]{Schootemeijer2018}. In other words, the two components of \starname~may be He-burning, classical WR stars rather than main-sequence WR stars. It is interesting to note that the morphology of the spectra resembles those of classical WR stars more than those of main-sequence WR stars \citep[see examples in][]{Crowther1997, Crowther1998}.

In this context, it is also  interesting to consider the less evolved "twin" of \starname, \object{Mk\,34} \citep{Tehrani2019}. \object{Mk\,34} was reported to host two stars of $\approx 140 + 130\,M_\odot$ (noting that these values rely on calibration to evolution models) that have roughly the same luminosities and temperatures as the components of \starname. However, they exhibit significantly lower mass-loss rates (smaller by a factor of three) and terminal velocities that are roughly twice as large compared with the components of \starname, which, empirically, are more typical for main sequence stars. If \starname~indeed comprises two classical WR stars, the components of \starname~may well be highly inflated, since they appear much larger than what one would expect ($R_* \approx 25\,R_\odot$ vs.\ $R_* \approx 1\,R_\odot$), as also observed for a multitude of Galactic classical WR stars \citep{Petrovic2006}.  Hence, \starname~could be a rare binary of two very massive, inflated classical WR stars, potentially the most massive classical WR stars ever weighed. Models by \citet{Graefener2021} suggest that such stars would have initial masses comparable to those obtained when assuming that the stars are homogeneous ($\approx 150\,M_\odot$ and $130\,M_\odot$), but their age would be roughly 3\,Myr instead of the 2\,Myr derived here.

\subsection{Future evolution}
\label{subsec:futev}

The similarity of both binary components and the fact that the system is still eccentric suggest that the stars have not interacted via mass-transfer in the past. On the other hand, given the evolutionary status of the system, the binary components of \starname~are also unlikely to interact in the future.  In fact, the final orbital period $P_{\rm f}$ between the two would only grow with time due to wind mass-loss as $P_{\rm f} / P_{\rm cur} \propto \left(M_{\rm tot, cur}/M_{\rm tot, f}\right)^2$, where $M_{\rm tot, cur}$ and $M_{\rm tot, f}$ are the total current and final mass of the system. With with an initial mass in excess of $140\,M_\odot$, the primary could be massive enough to undergo a pair-instability supernova, leaving no black hole behind \citep[e.g.,][]{Langer2007}. Assuming that the fate of both components is to end their lives as $\approx 20-30$~\msun\ BHs \citep{Belczynski2010}, we find that $P_{\rm f}  \approx 1-5\,$yr (depending on the current masses). That is, the system would become a BH+BH binary with an orbital period of a few years. 

At such large periods, interactions with tertiary companions become much more likely \citep[e.g.][]{Toonen2020}. With an integrated absolute V-band magnitude of $M_V \approx -8\,$mag, detecting any companions with current masses $\lesssim 50$~\msun\ (corresponding to light contributions smaller than $\approx 10$\%) via spectroscopy would be difficult, and it is therefore well possible that companions with masses as large as $\approx 50$~\msun\ reside in the vicinity of \starname.  A tertiary component could act to reduce the period of the binary through the Kozai-Lidov mechanism \citep{Lidov1962, Kozai1962}, eventually forcing the BHs to merge in a gravitational-wave event \citep[e.g.,][]{VignaGomez2021}. Alternatively, interactions with a third star could lead to instabilities that either eject the BHs at runaway velocities, or bind one of them to the tertiary star, forming a star+BH binary. Such objects could be the progenitors of high-mass X-ray binaries.

\subsection{The upper mass limit}
\label{subsec:uppermass}

Regardless of the evolutionary status of the system, our results imply initial masses of the order of $\approx 130-150\,M_\odot$ for the system (see Sect.\,\ref{subsec:evolution}).  The immediate conclusion is that stars as massive as 150~\msun\ form in the Local Group beyond reasonable doubt.  A similar conclusion was reached by \citet{Tehrani2019}, who reported initial  masses of $\approx 150\,M_\odot$ for the system \object{Mk\,34} -- the younger twin of \starname. The $150\,M_\odot$ limit is well within the $200-300\,M_\odot$ limit suggested by the bulk of the R136 population \citep{Schneider2018Sci, Graefener2021}. 

However, as discussed in Sect.\,\ref{sec:intro}, the derivation of masses in the range $200-300\,M_\odot$ relies on the assumption that the most massive stars  in the core of the Tarantula region -- \object{R\,136a1}, \object{R\,136a2}, \object{R\,136b}, and \object{R\,136c} --  are single \citep{Crowther2010, Bestenlehner2020}. \citet{Crowther2010} argued that lack of substantial X-ray emission in these objects indicates that they are single. However, as discussed in Sect.\,\ref{subsec:LCXR}, \starname~provides a clear counter example to this. Future effort should be dedicated into further investigating the multiplicity of the most massive stars, which have so-far been largely sensitive only to short-period ($P\lesssim 50\,$d) binaries \citep[e.g,][]{Schnurr2009}.

\subsection{Runaway status}
\label{subsec:runaway}

\begin{figure}
\centering
\includegraphics[width=0.5\textwidth]{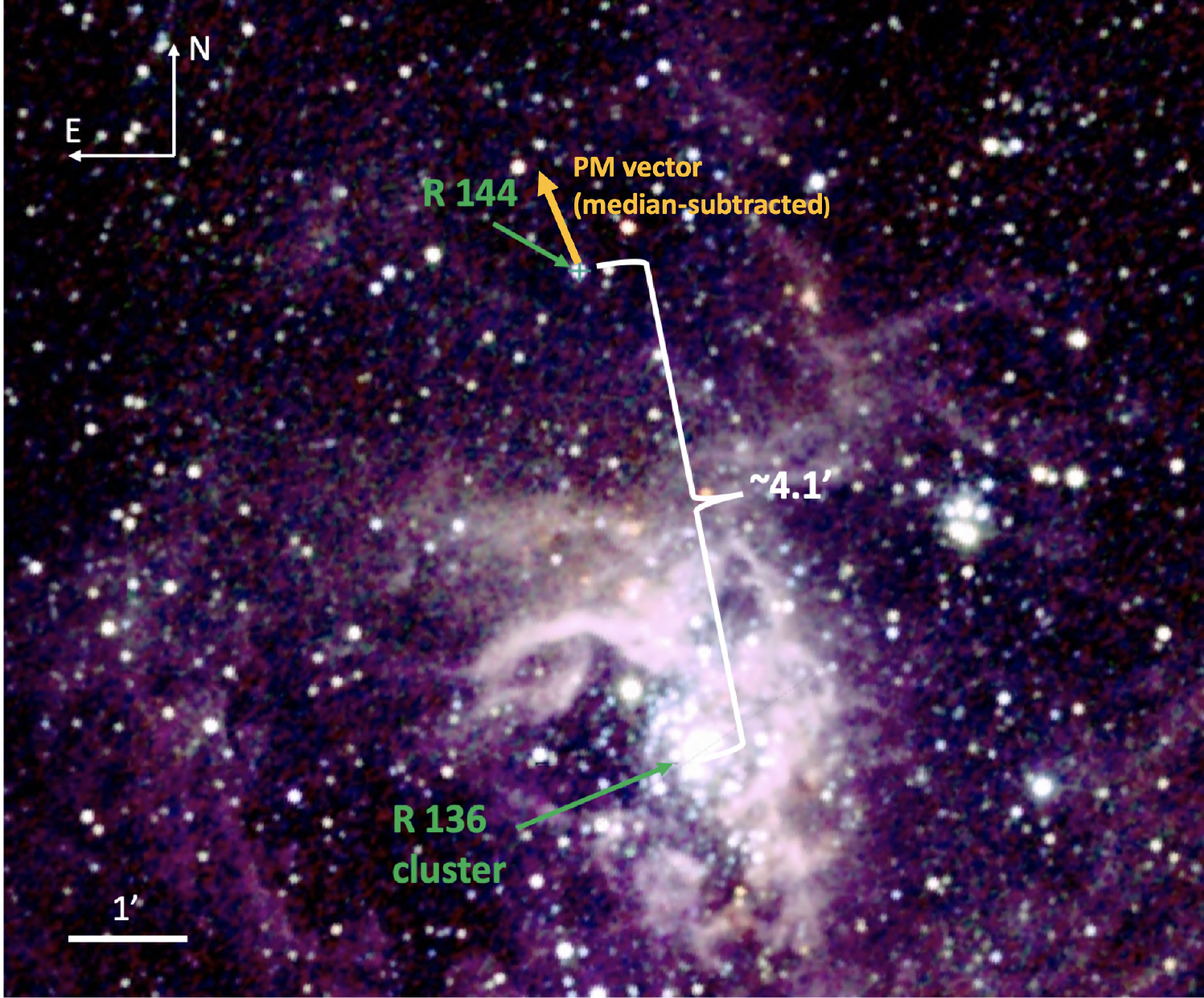}
\caption{2MASS JHK image \citep{Skrutskie2006}, highlighting the relative separation of \starname~from \object{R\,136} of about 4.1' ($\approx 60\,$pc). The median-subtracted proper motion arrow from Gaia eDR3 is also marked.
}
\label{fig:RW}
\end{figure}

\starname~is situated in relative isolation  with respect to the neighboring very massive and young cluster \object{R\,136}, roughly 4.1' apart (i.e. $\approx 60$\,pc,  Fig.\,\ref{fig:RW}). As discussed by \citet{Sana2013}, this could either mean that \starname~was formed in situ in relative isolation, or that it is a runaway ejected from a nearby cluster, most likely \object{R\,136}. The systemic velocity of $V_0 = 210\pm20$\kms~measured in our study puts it roughly $60\,$\kms~below the bulk of the Tarantula clusters \object{NGC 2070} and \object{R\,136} \citep[cf.][]{Almeida2017} of $\approx 270\pm10\,$\kms. While the measurement of absolute velocities of WR stars is model-dependent (Sect.\,\ref{subsec:RVmeas}), the observed offset is larger than $3\sigma$ when considering our conservative error estimate of $20\,$\kms. 

To further investigate the runaway status of \starname, we retrieved the proper motion (PM) components of \starname~and all stars within 4' from the Gaia eDR3 catalogue \citep{Gaia2020}. We computed the mean, median, and standard deviations of the PM components for all objects with a PM accuracy better than 0.1\,${\rm mas}\,{\rm yr}^{-1}$  in both PM components (464 objects in total). The standard deviation is $\overrightarrow{\sigma}_{\rm PM, 4'} = (0.16, 0.17)$\,${\rm mas}\,{\rm yr}^{-1}$ for the RA and DEC components. The difference between the median and average is negligible. We then subtracted the PM of \starname~from the median PM vector and obtain $\overrightarrow{ \Delta {\rm PM}}_{\rm R144} = (0.12 \pm 0.02, 0.28\pm0.03)$\,${\rm mas}\,{\rm yr}^{-1}$, where the errors are the Gaia eDR3 measurement errors of \starname. Evidently, the residual vector has a positive component pointing to the North-East (away from \object{R\,136}, Fig.\,\ref{fig:RW}). Within errors, the direction of the residual PM vector supports the idea that \starname~was ejected from \object{R\,136} at a significance larger than 3$\sigma$, with a transverse velocity component of $\approx 70\,$\kms.  If \starname~was ejected from \object{R\,136} in the past $\approx 2\,$Myr, a distance of 4.1' would require a runaway velocity $\varv \gtrsim 30\,$\kms. For a transverse velocity component of $\approx 70\,$\kms, the ejection needs to have happened roughly 1\,Myr ago. 

Assuming \starname~was indeed ejected from \object{R\,136}, the ejection was most likely initiated through dynamical interactions between \starname~and the massive \object{R\,136} cluster \citep{Fujii2011}. Unlike supernovae kicks, which tend to result in relatively modest ejection velocities \citep[e.g.,][]{Renzo2019}, dynamical interactions of very massive binaries with massive stars in the cluster can easily result in ejection velocities well above the escape velocity of the cluster.  While single stars are expected to be ejected first, repeated dynamical interactions over time continuously harden the "bully binary" and induce a net momentum to eject it from the cluster. However, it is worth noting that the dynamical age of the ejection is roughly half our derived age (2-3\,Myr), which is not easily explained in light of the lack of very massive stars of comparable ages in R\,136.  We encourage future investigations of this problem.

In conclusion, the systemic velocity of \starname~and its PM vector in combination with its relative isolation supports  the fact that it was ejected through dynamical interactions from the neighboring \object{R\,136} cluster, rather than having formed in situ.

% Following \citet{Fujii2011}, the squared ejection speed can be roughly estimated to be of the order of $G\,M_{\rm tot}\,{a}^{-1}$, where $G$ is the gravitational constant, $M_{\rm tot}$ the total mass of the binary, and $a$ the semi-major axis, which in the case of \starname~results in $\approx 400\,$\kms -- well above the necessary threshold to eject \starname. 

\section{Summary}
\label{sec:summary}

This study provides a comprehensive spectroscopic, photometric, and polarimetric analysis of the visually-brightest Wolf-Rayet star in the LMC, the binary \starname. We relied on multi-epoch ESO spectra acquired with X-SHOOTER and \mbox{(FLAMES-)UVES}, combined with FUSE and IUE spectra in the UV. We performed an orbital analysis and spectral disentangling of the system, and used the non-LTE Potsdam Wolf-Rayet (PoWR) code to derive the physical parameters of both components. Below, we summarise out conclusions:

\begin{itemize}
    \item \starname~is a binary comprising two WR stars (WN5/6h + WN6/7h) of nearly equal masses ($M_1 \sin^3 i = 48.3\pm1.8$ and $M_2 \sin^3 i = 45.5\pm1.8\,M_\odot$) orbiting their center of mass in an eccentric ($e=0.51$) orbit with a period of $P = 74.2$\,d. The primary is slightly hotter and more luminous than the secondary, and both stars exhibit substantial residual hydrogen ($X_{\rm H} \approx 0.4$) and nitrogen enrichment. Assuming the stars are still on the main sequence, their inferred age is $\approx 2\,$Myr.
    
    \item  The systemic velocity, proper-motion vector, and relative isolation of \starname~imply that it was ejected from the nearby \object{R\,136} cluster, presumably through dynamical interactions in the cluster.  
    
    \item  Contemporary evolution models that include enhanced mass-loss rates 
    \citep{Graefener2021} are successful in reproducing the low rotation and significant hydrogen depletion observed in both stars. 
    
    \item \starname~shows a periodically modulated light-curve that is well explained with a hybrid model invoking excess emission from wind-wind collisions and wind eclipses. The model enables us to derive the orbital inclination with a small formal error, $i=60.4\pm1.5^\circ$. In turn, this results in accurate dynamical masses of $74\pm4\,M_\odot$ and $69\pm4\,M_\odot$, making \starname~one of the only very massive binaries in the LMC for which an independent mass measurement is available. However, these masses are difficult to reconcile with the luminosities of $\log L = 6.44, 6.39\,[L_\odot]$  for the primary and secondary, which result in evolutionary masses of the order of $100-110\,M_\odot$. A good agreement between evolutionary and dynamical masses would be achieved for $i=50^\circ$, but this inclination does not seem to be consistent with the light-curve (however, see discussion in Appendix\,\ref{appendix:systematics}).
    
    \item  Taken at face value, the derived dynamical masses and luminosities imply that both components have classical Eddington factors of of $\Gamma_{\rm e} = 0.78 \pm0.10$, and are thus expected to be inflated due to their proximity to the Eddington limit. If the stars are on the main sequence, their derived radii imply that they are only slightly inflated. Alternatively it is possible that the components of \starname~are not on the main sequence, but are rather inflated classical (i.e., core He-burning) WR stars. If so, this would mean that \starname~potentially comprises the most massive classical (i.e., core He-burning) WR stars to have been weighed so far -- the evolved counterpart of the most massive binary thus weighed, \object{Mk\,34} \citep{Tehrani2019}.

\end{itemize}
 
\starname~is a one-of-a-kind  laboratory to study some of the most urgent questions at the upper-mass end, and poses several challenges to our theories of stellar evolution. To advance, we advocate for the acquisition of high-resolution spectra close to periastron passage, long-term high-precision photometry,  and phase-dependent polarimetry of \starname~and other very massive binaries in the Local Group.

\section*{Acknowledgments}
Dedicated in loving memory of Dr.\ Simon Clark, who passed away during preparation of this manuscript. We thank C.\ Evans for helpful discussions.  This research has made use of the VizieR catalogue access tool, CDS,
 Strasbourg, France (DOI : 10.26093/cds/vizier). The original description 
 of the VizieR service was published in 2000, A\&AS 143, 23. TS acknowledges funding from the European Research Council (ERC) under the European Union's Horizon 2020 research and innovation programme (grant agreement numbers 772225: MULTIPLES). TVR, DMB, and PM gratefully acknowledge support from the Research Foundation Flanders (FWO) by means of Junior and Senior Postdoctoral Fellowships, under contract No.~12ZB620N, No.~1286521N, and No.~12ZY520N, respectively. AFJM is grateful to NSERC (Canada) for financial aid.  FRNS has received funding from the European Research Council (ERC) under the European Union's Horizon 2020 research and innovation programme (Grant agreement No. 945806).  LM thanks the European Space Agency (ESA) and the Belgian Federal Science Policy Office (BELSPO) for their support in the framework of the PRODEX Programme. S.d.M.\ was funded in part by the European Union's Horizon 2020 research and innovation program from the European Research Council (ERC, Grant agreement No. 715063), and by the Netherlands Organization for Scientific Research (NWO) as part of the Vidi research program BinWaves with project number 639.042.728.

% \newpage
\begin{appendix}

\section{Observation log and RV measurements}
\label{appendix:log}

\begin{table*} 
\label{tab:log}
\small
\centering 
\caption{Observation log and measured radial velocities}
\begin{tabular}{ccccccc} \hline \hline
MJD & phase & Instrument & S/N  & $R$ & ${\rm RV}_1$ (\NVred) [km/s] & ${\rm RV}_2$ (\NIII) [km/s]  \\
 &  &  &  &  & \kms & \kms  \\
\hline
52956.80 & 0.41 & UVES & 100 & 55000  & 196 $\pm$ 5 & 207 $\pm$ 6\\ 
52958.82 & 0.44 & UVES & 150 & 55000  & 179 $\pm$ 5 & 212 $\pm$ 6\\ 
55308.98 & 0.11 & X-SHOOTER & 230 & 9700  & 331 $\pm$ 21 & -\\ 
55452.39 & 0.04 & X-SHOOTER & 130 & 9700  & 356 $\pm$ 19 & -3 $\pm$ 22\\ 
55452.40 & 0.04 & X-SHOOTER & 210 & 9700  & 373 $\pm$ 21 & 18 $\pm$ 24\\ 
55580.08 & 0.76 & X-SHOOTER & 480 & 9700  & 87 $\pm$ 17 & 266 $\pm$ 21\\ 
55580.09 & 0.77 & X-SHOOTER & 160 & 9700  & 81 $\pm$ 19 & 255 $\pm$ 21\\ 
55585.15 & 0.83 & X-SHOOTER & 480 & 9700  & 96 $\pm$ 19 & 274 $\pm$ 24\\ 
55589.18 & 0.89 & X-SHOOTER & 350 & 9700  & 103 $\pm$ 20 & 302 $\pm$ 25\\ 
55604.03 & 0.09 & X-SHOOTER & 190 & 9700  & 357 $\pm$ 22 & 21 $\pm$ 24\\ 
55672.98 & 0.02 & X-SHOOTER & 170 & 9700  & 334 $\pm$ 19 & 42 $\pm$ 22\\ 
55674.48 & 0.04 & X-SHOOTER & 180 & 9700  & 327 $\pm$ 23 & 69 $\pm$ 25\\ 
55697.95 & 0.35 & FEROS & 60 & 48000  & 209 $\pm$ 10 & 163 $\pm$ 9\\ 
56210.88 & 0.27 & FLAMES-UVES & 110 & 47000  & 224 $\pm$ 6 & 146 $\pm$ 5\\ 
56217.85 & 0.36 & FLAMES-UVES & 70 & 47000  & 207 $\pm$ 6 & 181 $\pm$ 6\\ 
56243.86 & 0.71 & FLAMES-UVES & 100 & 47000  & 105 $\pm$ 6 & 296 $\pm$ 6\\ 
56256.78 & 0.88 & FLAMES-UVES & 90 & 47000  & 111 $\pm$ 7 & 273 $\pm$ 6\\ 
56257.65 & 0.90 & FLAMES-UVES & 90 & 47000  & 117 $\pm$ 7 & 291 $\pm$ 6\\ 
56277.83 & 0.17 & FLAMES-UVES & 130 & 47000  & 294 $\pm$ 6 & 84 $\pm$ 5\\ 
56283.57 & 0.25 & FLAMES-UVES & 70 & 47000  & 238 $\pm$ 7 & 139 $\pm$ 6\\ 
56294.72 & 0.40 & FLAMES-UVES & 90 & 47000  & 193 $\pm$ 6 & 204 $\pm$ 6\\ 
56295.70 & 0.41 & FLAMES-UVES & 110 & 47000  & 193 $\pm$ 5 & 210 $\pm$ 6\\ 
56305.75 & 0.54 & FLAMES-UVES & 100 & 47000  & 158 $\pm$ 6 & 243 $\pm$ 6\\ 
56306.74 & 0.56 & FLAMES-UVES & 120 & 47000  & 155 $\pm$ 17 & 234 $\pm$ 6\\ 
56316.73 & 0.69 & FLAMES-UVES & 130 & 47000  & 112 $\pm$ 7 & 272 $\pm$ 5\\ 
56338.02 & 0.98 & X-SHOOTER & 160 & 9700  & 224 $\pm$ 19 & 126 $\pm$ 22\\ 
56349.54 & 0.13 & FLAMES-UVES & 100 & 47000  & 308 $\pm$ 6 & 64 $\pm$ 6\\ 
56352.55 & 0.17 & FLAMES-UVES & 80 & 47000  & 289 $\pm$ 7 & 104 $\pm$ 6\\ 
56356.53 & 0.23 & FLAMES-UVES & 110 & 47000  & 258 $\pm$ 7 & 123 $\pm$ 6\\ 
56571.87 & 0.13 & FLAMES-UVES & 40 & 47000  & 343 $\pm$ 13 & 72 $\pm$ 22\\ 
56582.86 & 0.28 & FLAMES-UVES & 10 & 47000  & 212 $\pm$ 22 & 143 $\pm$ 24\\ 
56586.77 & 0.33 & FLAMES-UVES & 70 & 47000  & 200 $\pm$ 17 & 190 $\pm$ 6\\ 
56597.75 & 0.48 & FLAMES-UVES & 30 & 47000  & 143 $\pm$ 12 & 239 $\pm$ 7\\ 
56620.78 & 0.79 & FLAMES-UVES & 80 & 47000  & 105 $\pm$ 12 & 290 $\pm$ 6\\ 
56627.69 & 0.88 & FLAMES-UVES & 40 & 47000  & 100 $\pm$ 17 & 248 $\pm$ 15\\ 
56645.57 & 0.12 & FLAMES-UVES & 80 & 47000  & 325 $\pm$ 17 & 38 $\pm$ 20\\ 
56653.80 & 0.23 & FLAMES-UVES & 80 & 47000  & 231 $\pm$ 42 & 131 $\pm$ 6\\ 
56693.63 & 0.77 & FLAMES-UVES & 60 & 47000  & 95 $\pm$ 10 & 306 $\pm$ 6\\ 
56697.69 & 0.83 & FLAMES-UVES & 30 & 47000  & 81 $\pm$ 19 & 266 $\pm$ 10\\ 
56703.66 & 0.91 & FLAMES-UVES & 70 & 47000  & 111 $\pm$ 12 & 273 $\pm$ 6\\ 
56705.04 & 0.92 & X-SHOOTER & 520 & 9700  & 141 $\pm$ 17 & 231 $\pm$ 21\\ 
56705.05 & 0.92 & X-SHOOTER & 220 & 9700  & 147 $\pm$ 18 & 232 $\pm$ 21\\ 
56706.05 & 0.94 & X-SHOOTER & 310 & 9700  & 166 $\pm$ 18 & 209 $\pm$ 22\\ 
56707.06 & 0.95 & X-SHOOTER & 220 & 9700  & 173 $\pm$ 18 & 193 $\pm$ 22\\ 
56708.04 & 0.96 & X-SHOOTER & 260 & 9700  & 196 $\pm$ 19 & 188 $\pm$ 21\\ 
56709.04 & 0.98 & X-SHOOTER & 810 & 9700  & 215 $\pm$ 18 & 152 $\pm$ 22\\ 
56710.03 & 0.99 & X-SHOOTER & 120 & 9700  & 249 $\pm$ 18 & 107 $\pm$ 22\\ 
56711.02 & 0.01 & X-SHOOTER & 120 & 9700  & 282 $\pm$ 20 & 74 $\pm$ 22\\ 
56712.02 & 0.02 & X-SHOOTER & 160 & 9700  & 331 $\pm$ 17 & 41 $\pm$ 22\\ 
56713.03 & 0.03 & X-SHOOTER & 380 & 9700  & 345 $\pm$ 18 & 18 $\pm$ 22\\ 
56714.03 & 0.05 & X-SHOOTER & 420 & 9700  & 345 $\pm$ 8 & 23 $\pm$ 22\\ 
56714.55 & 0.05 & FLAMES-UVES & 50 & 47000  & 346 $\pm$ 13 & 30 $\pm$ 8\\ 
56715.03 & 0.06 & X-SHOOTER & 270 & 9700  & 349 $\pm$ 18 & 36 $\pm$ 22\\ 
56716.04 & 0.07 & X-SHOOTER & 190 & 9700  & 338 $\pm$ 18 & 40 $\pm$ 21\\ 
56719.54 & 0.12 & FLAMES-UVES & 80 & 47000  & 344 $\pm$ 9 & 79 $\pm$ 6\\ 
56723.70 & 0.18 & FLAMES-UVES & 40 & 47000  & 302 $\pm$ 20 & -\\ 
\hline
\end{tabular}
\end{table*}

\section{TESS data}
\label{appendix:TESS}

\begin{figure}
\centering
\includegraphics[width=0.5\textwidth]{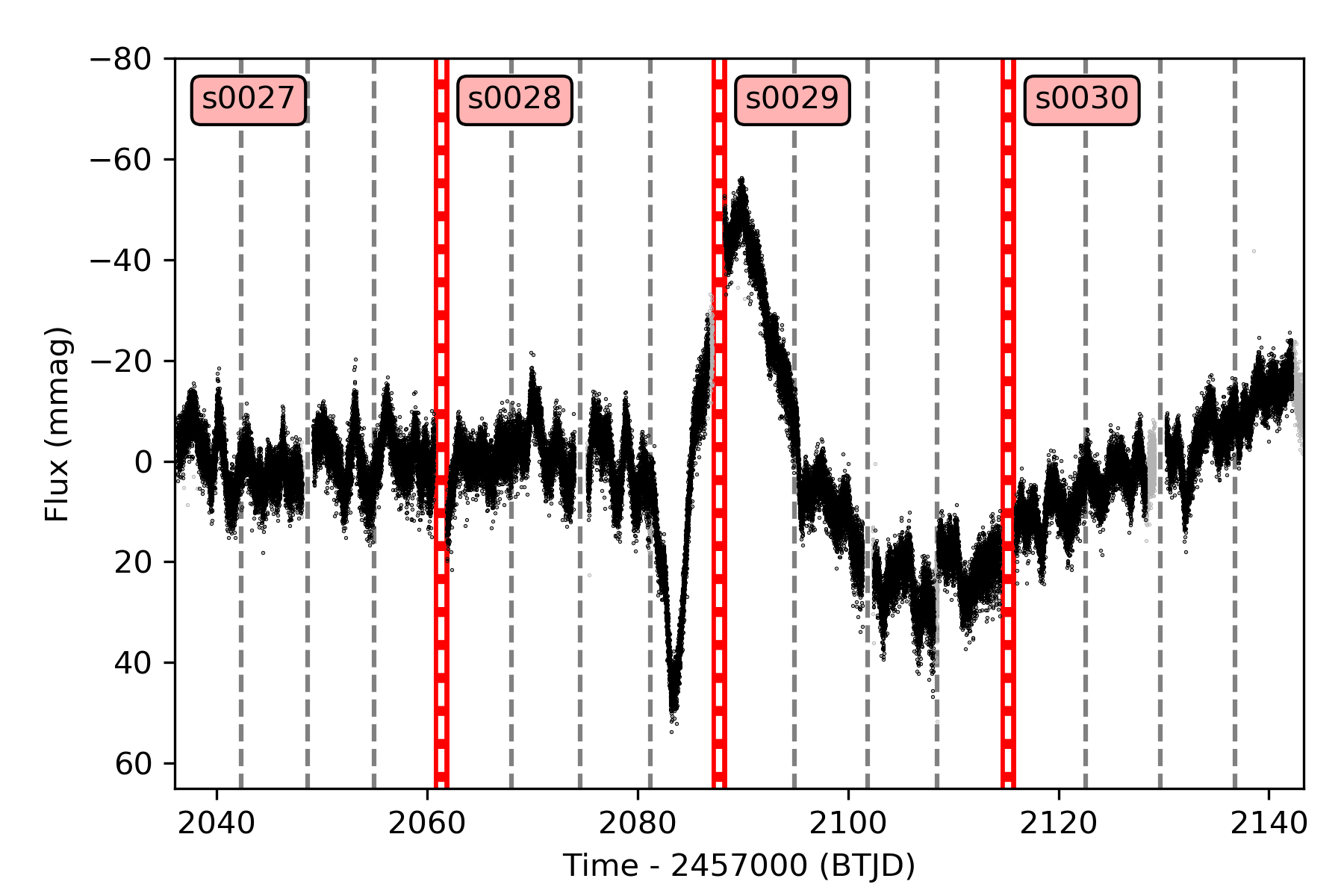}
\caption{The short-cadence TESS light-curve (black dots), obtained for sectors 27 to 30. Flagged (less reliable) data points are marked in light~gray, and the dashed gray vertical lines indicate the time stamps of the TESS angular momentum dumps.
}
\label{fig:TESSJD}
\end{figure}

Figure\,\ref{fig:TESSJD} shows the entire TESS photometry, which comprises four different sectors that are stitched together. The data as a whole covers roughly 1.5 orbital cycles and one periastron passage. With a pixel size of 21'', contamination with nearby sources is possible, though comparison with the GAIA early DR3 catalogue \citep{Gaia2018, Gaia2020} suggests the contamination does not exceed 1-2\%.   The high-precision TESS photometry data reveal interesting substructures in the light-curve indicative of pulsational activity, though the time coverage does not enable a reliable period analysis of these structure to identify their possible origin.

As Fig.\,\ref{fig:TESSJD} suggests, combining the TESS data with the ASAS-SN is not trivial, as the stitching and calibration of the individual sectors can lead to large systematic errors. In fact, as Fig.\,\ref{fig:TESSJD} readily shows, the long-term increase seen in the ASAS-SN data between $\phi = 0.2$ and $0.8$ is seen only to the right of periastron, but not to the left, implying a substantial systematic error in the stitched light-curve.

\begin{figure}
\centering
\includegraphics[width=0.5\textwidth]{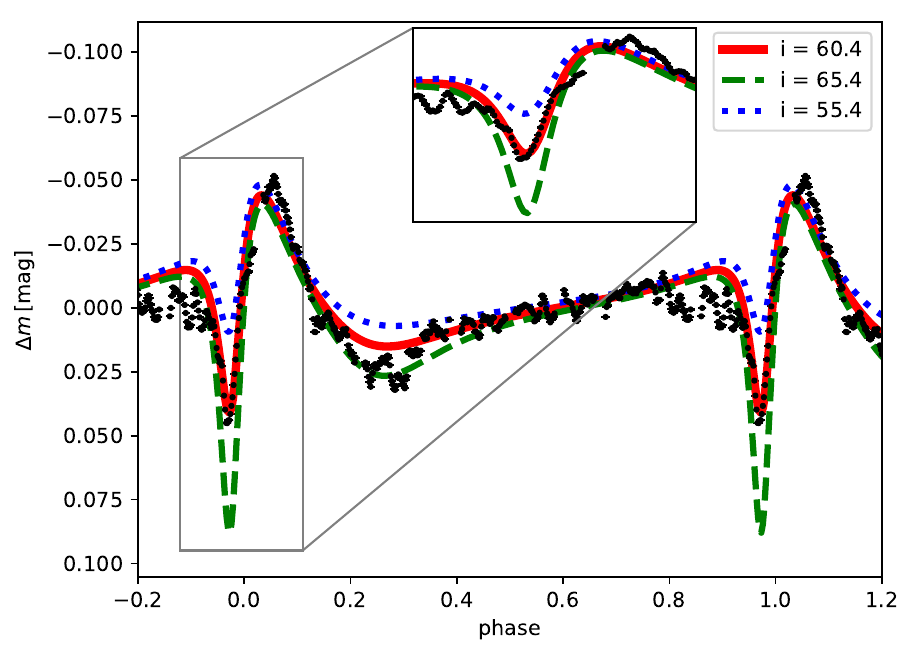}
\caption{As Fig.\,\ref{fig:hybridModelInc}, but comparing with the phased TESS data instead of the ASAS-SN data.
}
\label{fig:TESSinc}
\end{figure}

Because of these systematics, and since the TESS data only cover 1.5 orbital cycles and are hence greatly affected by stochasticity in the system, we do not use the TESS data in our model fitting. However, we do use it to refine the orbital period of the system, as described in Sect.\,\ref{subsec:LC}. Nevertheless, we present in Fig.\,\ref{fig:TESSinc} a comparison between the phased TESS light-curve and our hybrid light-curve model shown in Sect.\,\ref{subsec:LC}. There are some apparent deviations, but overall the model qualitatively well reproduces the light-curve. The TESS data seem to suggest a deeper secondary minimum at $\phi = 0.28$ (which implies either even higher inclinations or an increased mass-loss rate for the secondary), but with only 1.5 cycles, we avoid an over-interpretation of this mismatch.

\section{PHOEBE modelling of the light-curve}
\label{appendix:PHOEBE}

\begin{figure}
\centering
\includegraphics[width=0.5\textwidth]{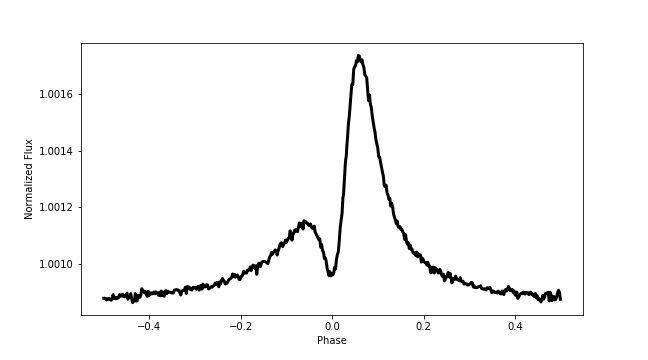}
\caption{Plotted is our PHOEBE model of \starname, using the parameters given in Table\,\ref{tab:parameters} as input. Due to the different scale of amplitude of the data and model, the data are not plotted for clarity (cf.\ Fig.\,\ref{fig:ASASLC}).
}
\label{fig:PHOEBEfitSmall}
\end{figure}

We attempted to model the ASAS-SN light-curve using the Physics of Eclipsing Binaries (PHOEBE) light-curve modelling tool \citep{Prsa2016}. Starting from the parameters derived using the radial velocity curves and spectroscopic fitting we endeavoured to derive a self consistent model. However, our efforts led to the model seen in Fig\,\ref{fig:PHOEBEfitSmall}. While the "heartbeat" shape is recovered, the amplitude  is nearly two orders of magnitude lower than that seen in the actual light-curve. Moreover, the minimum is obtained at periastron, while the observed minimum is at $\phi=-0.026$ (inferior conjunction), suggesting that tidal distortion is not the dominant mechanism governing the shape of the light-curve.

\section{Systematic uncertainties in the light-curve model}
\label{appendix:systematics}

\setlength{\tabcolsep}{3pt}
\begin{table*}
\centering
    \caption{Inferred parameters from our hybrid light-curve model (2829 fitting points, combined RV + light-curve).}
    \begin{tabular}{ccccccccc} \hline \hline
    model    & $\chi^2$\,$^{\rm a}$ &  $A_{\rm WWC}$ & $\gamma_{\rm WWC}$ & $\dot{M}_1$ [\smy] & $\dot{M}_2$ [\smy] &  $i$ [deg] & $f_1/f_2(V)$  & $\beta$  \\
        \hline
        {\bf adopted}    & 1.0000 (2827) &  $0.12 \pm 0.01$ & $1.5 \pm0.1$ & $-4.29 \pm 0.05$ & $-4.42 \pm 0.05$ & $60.4 \pm 1.5$ & 0.79 (fixed) & 1 (fixed)  \\ 
      \hline        
      A2    &   1.0000 (2827)   &  $0.12 \pm 0.01$ & $1.5 \pm0.1$ & $-4.13 \pm 0.05$ & $-4.56 \pm 0.05$ & $60.4 \pm 1.5$ & {\bf 1.58 (fixed)} & 1 (fixed) \\ 
      A3    &  1.0000 (2827)  &  $0.12 \pm 0.01$ & $1.5 \pm0.1$ & $-4.39 \pm 0.05$ & $-4.22 \pm 0.05$ &  $60.4 \pm 1.5$  & {\bf 0.40 (fixed)}  & 1 (fixed) \\   
      A4    & 1.0000 (2827) &  $0.12 \pm 0.01$ & $1.5\pm0.1$ & {\bf -4.38 (fixed)} & {\bf -4.34 (fixed)} & $61.4\pm0.3$ & $0.53\pm0.02$ & 1 (fixed) \\  
    %   A5    &   1.0371 &  $0.12 \pm 0.01$ & {\bf 1 (fixed)} & $-4.35 \pm 0.05$ & $-4.43 \pm 0.05$ & $61.0\pm1.5$ & 0.79 (fixed)  & 1 (fixed) \\              
      A5    &  1.0112 (2860) &  $0.19 \pm 0.01$ & $1.3 \pm0.1$ & $-3.95 \pm 0.01$ & $-4.08 \pm 0.01$ & {\bf 50 (fixed)} & 0.79 (fixed)  & 1 (fixed) \\ 
      A6    &  1.0000 (2827) &  $0.11 \pm 0.01$ & $1.5 \pm0.1$ & $-4.34 \pm 0.05$ & $-4.44 \pm 0.04$ & $59.4\pm1.5$ & 0.75 (fixed)  & {\bf 2 (fixed)} \\         
        % 0.69601225 &  $0.15 \pm 0.02$ & $1.6 \pm0.1$ & $-4.32 \pm 0.05$ & $-4.41 \pm 0.05$ & $58.9 \pm 1.6$ & 0.754 (fixed)  & $-0.004 \pm 0.003$ \\         
    \hline
    \end{tabular}
\tablefoot{$^{\rm a}$ reduced and non-reduced $\chi^2$ values with 12 degrees of freedom and 2839 fitting points. Since the photometry errors appear to be overestimated they were reduced by a factor of 0.82 to obtain $\chi^2$ values of the order of unity (see Sect.\,\ref{subsec:LC}).
}    
\label{tab:LCmodel}
\end{table*}

Modelling of the ASAS-SN data enables an accurate derivation of the orbital inclination of $i = 60.4\pm1.5^\circ$. However, the error reflects the formal fitting error, and does not account for any potential systematic errors originating in the assumptions in the model. Our numerical attempts to vary different assumptions are shown in Table\,\ref{tab:LCmodel}. Below, we summarise our results in the context of their impact on the derived value of $i$, and discuss additional sources of uncertainty and their potential impact.

\begin{enumerate}
    \item \emph{Altering the light ratio:} we explored the impact of fixing the light ratio to twice and half the adopted value. This has a substantial impact on the mass-loss rates, but has no impact on the derived inclination. The fit quality ($\chi^2$) remains unchanged, which is anticipated as $\dot{M}_1$, $\dot{M}_2$ and $f_1/f_2$ are degenerate.
    \item {Altering $\dot{M_1}, \dot{M_2}$:} similarly, we fixed $\dot{M}$ to the values derived spectroscopically. This only leads to a slight increase (by $1^\circ$) of the orbital inclination, to compensate for the net decrease of density in the wind of the primary. The fit quality is comparable.
    \item \emph{Altering $\beta$}: The $\beta$-exponent in the velocity law determines the steepness of acceleration in the wind. Larger $\beta$ values result in enhanced wind densities, and could hence result it higher inclination angles. \citet{Lamontagne1996} derived analytical equations for $\beta = 0$ and $\beta = 1$. However, larger $\beta$ values will result in denser winds, which in turn should result in lower inclinations for the same column density.  we derive the relevant equations for $\beta = 2$ (Appendix\,\ref{appendix:beta}) to explore the impact of increasing $\beta$. While this indeed yields a slightly lower inclination, the decrease is small (by $1^\circ$) and still falls within the formal error margin. The fit quality is comparable, suggesting that $\beta$ and $i$ are degenerate parameters. However, $\beta$ cannot be much larger than $\beta=2$ based on our spectroscopic analysis (Sect.\,\ref{subsec:nonLTE}). 
    \item \emph{Varying $\alpha$, $R_*$, or $v_\infty$:} We have altered the number of electrons per baryon $\alpha$, the stellar radii $R_*$, and the terminal velocities $v_\infty$ within formal errors. Some impact, though a small one ($\approx 0.05\,$dex) could be seen on the mass-loss rates, but no notable change was obtained in the value of $i$. 
    \item \emph{Ionization structure in the wind:} The model relies on a constant $\alpha$, assuming that H and He are fully ionized in the wind. Our atmosphere models reveal that this assumption is valid for the hotter primary in the relevant domain, but breaks in the cooler secondary a few stellar radii above the surface, introducing a potential depth-dependence in $\alpha$. However, as $\alpha$ is primarily a multiplicative parameter determining the strength of absorption, any changes in $\alpha$ will at first order impact the derived mass-loss rates from this method, with impact on the inclination being second-order. 
    \item \emph{Impact of WWC cone:} At the region where the two winds collide, substantial density enhancements are expected. Moreover, the cone slices the stellar winds, thereby altering the optical depth across the line-of-sight in a non-trivial manner. It is difficult to assess the impact of these deviations from spherical symmetry without a full 3D radiative transfer model. However, the impact of this is expected to be very different at inferior and superior conjunction. The fact that our model shows no notable deviations during both eclipses suggests that the impact of the WWC cone in this context cannot be substantial. 
\end{enumerate}

It is beyond the scope of this paper to construct a full numerical hydrodynamic/radiative-transfer model to further investigate additional sources of uncertainty (e.g., depth-dependent ionization, impact of density enhancement in WWC cone, finite sizes of stellar disks). Our tests imply that the derived value of $i$ is robust within the provided errors, but accurate modelling of the problem in future works is encouraged.

\section{Wind-eclipse model with $\beta=2$}
\label{appendix:beta}
We provide an extension of the wind-eclipse model of \citet{Lamontagne1996}  for the case of a $\beta$-law velocity exponent of $\beta=2$. To compute the optical depth from the wind one needs to evaluate the integral
\begin{eqnarray}
\tau = k\int_{-\epsilon}^{\infty}\frac{d(z/\mathcal{D})}{(r/\mathcal{})^2(1-R_*/r)^\beta},
\end{eqnarray}
where
\begin{eqnarray}
\begin{aligned}
\epsilon = \sin i \cos \varpi, \quad \left(\frac{r}{\mathcal{D}}\right)^2=\cos^2 i \cos^2\varpi + \sin^2 \varpi+\left(\frac{z}{\mathcal{D}}\right)^2,
\end{aligned}
\end{eqnarray}
$\mathcal{D}$ is the instantaneous separation between the two components, $z$ is the Cartesian coordinate towards the observer, $r$ is the radial distance from the center of the eclipsing star (see figure\,1 in \citealt{Lamontagne1996}), and $k$ is the Thomson opacity.

Defining $\gamma=\sqrt{\cos^2 i \cos^2\varpi + \sin^2 \varpi}$, $\psi=\sqrt{\gamma^2-\delta^2}$ and $\delta=R_*/a$, the analytical solution to the integral for the case $\beta=2$ is given by
\begin{eqnarray}
\begin{aligned}
\frac{\tau}{k}=&\frac{\delta}{\psi^2}\left(1+\frac{\epsilon \sqrt{\epsilon^2+\gamma^2}+\delta^2\epsilon^2}{\psi^2+\epsilon^2}\right) \\
&+\frac{\gamma^2}{\psi^3}\left(\arcsin\left(\frac{\delta}{\gamma}\right)+\frac{|\epsilon|}{\epsilon}\arcsin\left(\sqrt{\frac{\delta^2\epsilon^2}{\gamma^2(\psi^2+\epsilon^2)}}\right)\right)\\
&+\frac{\pi \gamma^2}{2\psi^3}\left(\frac{\pi}{2}+\arctan\left(\frac{\epsilon}{\psi}\right)\right).
\end{aligned}
\end{eqnarray}

\end{appendix}

% \begin{figure}
% \centering
% \includegraphics[width=0.5\textwidth]{He_I_4472.pdf}
% \caption{As Fig.\,\ref{fig:NIVdis}, but for the N\,{\sc v}\,$\lambda 4603$. The analysis is performed using the FLAMES-UVES data.  The minimum is at $K_1 = 136 \pm 4\,$\kms, $K_2 = 143\pm 4$\,\kms.
% }
% \label{fig:HeIWWC}
% \end{figure}

%FFFFFFFFFFFFFFFFFFFFFFFFFFFFFFFFFFFFFFFFFFFFFFFFFFFFFFFFFFFFFF
%FFFFFFFFFFFFFFFFFFFFFFFFFFFFFFFFFFFFFFFFFFFFFFFFFFFFFFFFFFFFFF

\bibliographystyle{aa}
\bibliography{papers}

\end{document}